# The Solar Neighborhood L: Spectroscopic Discovery of K Dwarfs Younger Than 1 Gyr and New Binaries within 30 pc

Hodari-Sadiki Hubbard-James[1,2,4], D. Xavier Lesley[2,3], Todd J. Henry[2,4], Leonardo A. Paredes[1,2,4], and Azmain H. Nisak[1,4]
[1] Department of Physics and Astronomy, Georgia State University, Atlanta, GA 30302, USA; hjames12@gsu.edu
[2] RECONS Institute, Chambersburg, PA 17201, USA
[3] Department of Physics, Southern Connecticut State University, New Haven, CT 06515, USA

## Abstract

As part of a comprehensive effort to characterize the nearest stars, the CHIRON echelle spectrograph on the CTIO/SMARTS 1.5 m telescope is being used to acquire high-resolution ($R = 80,000$) spectra of K dwarfs within 50 pc. This paper provides spectral details about 35 K dwarfs from five benchmark sets with estimated ages spanning 20 Myr–5.7 Gyr. Four spectral age and activity indicators are tested, three of which aligned with the estimated ages of the benchmark groups—the Na I doublet (5889.95 and 5895.92 Å), the H$\alpha$ line (6562.8 Å), and the Li I resonance line (6707.8 Å). The benchmark stars are then used to evaluate seven field K dwarfs exhibiting variable radial velocities for which initial CHIRON data did not show obvious companions. Two of these stars are estimated to be younger than 700 Myr, while one exhibits stellar activity unusual for older K-dwarf field stars and is possibly young. The four remaining stars turn out to be spectroscopic binaries, two of which are being reported here for the first time with orbital periods found using CHIRON data. Spectral analysis of the combined sample of 42 benchmark and variable radial velocity stars indicates temperatures ranging from 3900 to 5300 K and metallicities from $-0.4 < $ [Fe/H] $ < +0.2$. We also determine $\log g = 4.5$–4.7 for main-sequence K dwarfs. Ultimately, this study will target several thousand of the nearest K dwarfs and provide results that will serve present and future studies of stellar astrophysics and exoplanet habitability.

*Unified Astronomy Thesaurus concepts:* K dwarf stars (876); Late-type dwarf stars (906); Solar neighbourhood (1509); Spectroscopy (1558); Stellar activity (1580); Stellar ages (1581); Stellar associations (1582); Stellar properties (1624)

## 1. Introduction

A key sample of stars to be explored for other planetary systems includes the Sun's slightly less massive, cooler cousins known as K dwarfs. In the solar neighborhood, K dwarfs account for ∼12% of all stars (Henry et al. 2006; with updates at www.recons.org), and their longevity makes them ideal hosts for habitable planets. A critical aspect in evaluating an individual planet's environment is the age of the star it orbits, i.e., is the star young, adolescent, or mature? For example, stellar youth is often associated with elevated ultraviolet luminosity, chromospheric flares, and a general increase in stellar activity, all factors linked to the magnetic dynamo at the star's core (Davenport et al. 2019). Skumanich (1972) discussed the link between stellar age and activity, describing how magnetic breaking would reduce the rotation speed of a star and dampen its dynamo, causing flares to weaken (Skumanich 1986). Consequently, young and active host stars are unlikely to provide stable environments and present challenges to astronomers attempting to estimate the locations of habitable realms for orbiting planets (Segura et al. 2010). Instead, post-adolescent hosts that have low levels of stellar activity (Luger et al. 2015) are preferred, given that they provide steadier, more durable conditions.

Our study aims to provide statistics on the youth and maturity of the nearby K-dwarf population, here defined to be stars within 50 pc. This is part of a larger REsearch Consortium On Nearby Stars (RECONS; www.recons.org) effort to characterize the closest ∼5000 K dwarfs. These stars are arguably among the best of all stars to host habitable planets (Cuntz & Guinan 2016), and our work will provide a list of key targets for exoplanet surveys from the ground, as well as for both current and future exoplanet atmosphere studies from space-based platforms, e.g., NASA's Hubble Space Telescope (HST) and James Webb Space Telescope (JWST), as well as the European Space Agency's upcoming Twinkle mission (Mollière et al. 2017; Edwards et al. 2019).

Currently, the only star with a precise and fundamental stellar age measurement is the Sun, determined via calculations related to the decay of radioactive isotopes in meteorites (Soderblom et al. 2014). All other age estimates are less accurate and come from measuring semi-fundamental properties of stars, or are model derived and based on other scientific inferences (Soderblom 2010; Soderblom et al. 2014). Popular techniques used to estimate ages for G-, K-, and M-type dwarfs include lithium depletion, utilized in Skumanich (1972), White et al. (2007), López-Santiago et al. (2010), and Binks & Jeffries (2014); kinematic motions in the Galaxy, highlighted in López-Santiago et al. (2010) and Mamajek & Bell (2014); and gyrochronology or stellar rotation studies, such as those of Brandt & Huang (2015), Gossage et al. (2018), and Skumanich (1972). These methods are often combined, in order to improve the reliability of the age estimates (Soderblom 2010).

---
[4] Visiting Astronomer, Cerro Tololo Inter-American Observatory. CTIO is operated by AURA, Inc., under contract to the National Science Foundation.

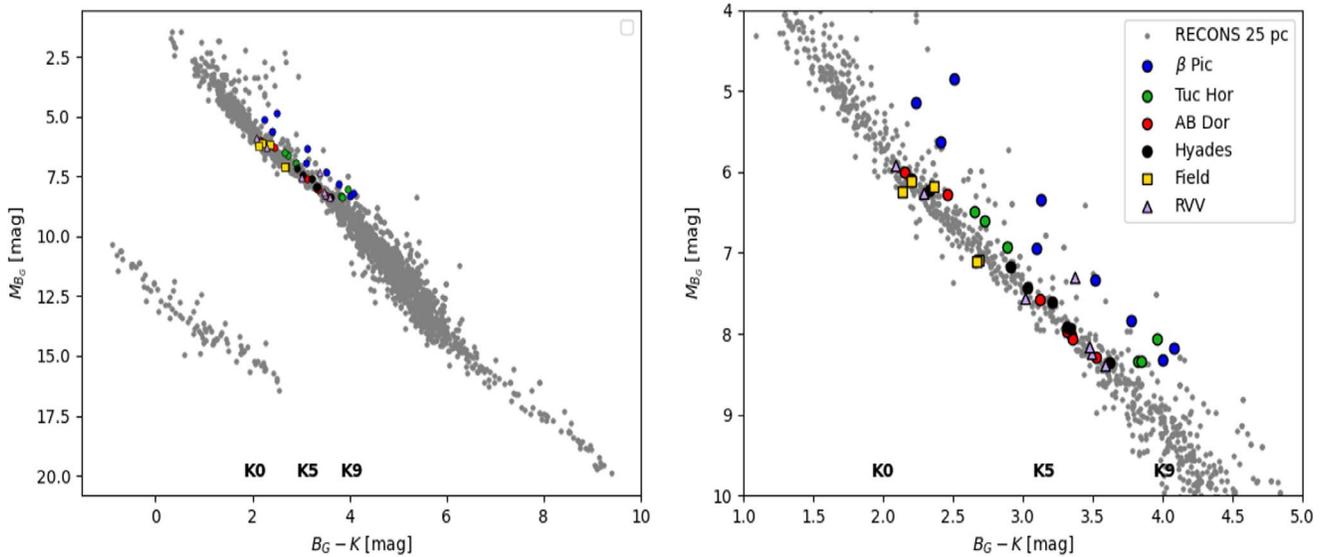

**Figure 1.** Left: H-R diagram highlighting the members of the benchmark sample targeted to map spectral features related to age and activity for K dwarfs. Points are color-coded as in the legend for $\beta$ Pic, Tuc-Hor, AB Dor, Hyades, field K dwarfs, and RV variable dwarfs (RVV). Smaller gray points represent stars in the RECONS 25 pc sample. Right: zoomed-in view to focus on the stars investigated in this paper, with the same background stars present. Magnitude information is obtained from Gaia EDR3 ($BP = B_G$) and Two Micron All Sky Survey (2MASS; $K_s = K$). Absolute $B_G$ magnitudes were derived using Gaia EDR3 parallax measurements. Guidelines for spectral types are given in both panels.

In addition to age, other stellar parameters, such as effective temperature, metallicity, surface gravity, and rotational velocity, come into play when estimating a planetary system's potential habitability. For example, an accurate measurement of a star's effective temperature is a prerequisite to setting the boundaries of the so-called "Goldilocks Zone," while metallicity may be an indicator of the amount of planet-building material available in the system. A low surface gravity points to a star that either has not yet reached the main sequence or has evolved off of it, while estimating a star's rotational velocity gives insight into the stellar activity affecting any surrounding planetary system (Cuntz & Guinan 2016; Brandt & Huang 2015). Overall, the assessment of any particular star as a host of an orbiting planet requires observations that can be used to evaluate vital stellar parameters like temperature, supplemented with additional information related to age, activity, and multiplicity.

In this work we investigate the utility of spectral features at optical wavelengths observed in K dwarfs that are potentially linked to age—the Na I doublet at 5890/5896 Å, the H$\alpha$ line at 6563 Å, the Li I resonance line at 6708 Å, and one of the Ca II infrared triplet lines at 8542 Å—to create a rubric for estimating the ages and activity levels of these stars. We also utilize the Empirical SpecMatch code of Yee et al. (2017) to derive stellar parameters through spectral modeling. To accomplish these tasks, we use high-resolution spectra ($R = 80,000$) from the Small and Moderate Aperture Research Telescope System (SMARTS) 1.5 m telescope and CHIRON echelle spectrograph for 35 K dwarfs in five groups that have age estimates. Our rubric is then tested on an initial set of seven nearby K dwarfs that exhibited variable radial velocities (RVs) in early CHIRON data sets. These stars may be particularly young or active, or the variations could be caused by companions. Stellar parameters, including line equivalent widths (EWs), temperatures, metallicities, surface gravities, and rotational velocities, are evaluated for all 42 stars analyzed in this study. As an ensemble, these stars constitute a benchmark sample of young stars of known age, a set of nearby K dwarfs of moderate ages determined using isochrones, and a number of new nearby young star candidates, all observed methodically with the same instrument and with parameters derived uniformly.

## 2. Sample Selection

There are two samples of K dwarfs targeted in this effort—a primary sample of (presumably) field stars and a benchmark comparison sample of stars with estimated ages. The Hertzsprung–Russell (H-R) diagrams presented in Figure 1 (left panel is an overview of stars within 25 pc; right panel is a zoom-in of the K-dwarf region) show that the majority of field K dwarfs and those from older groups are located on the main sequence, whereas the younger stars in the benchmark sample are sometimes located above the main sequence. Here we define K dwarfs as having Gaia EDR3 absolute BP (hereafter $B_G$) magnitudes of $M_{B_G} = 5.0 - 8.5$ (equivalent to 5.8–8.8 in absolute Johnson $V$ photometry) and colors of $B_G$–$K$ = 2.0–4.0 (1.9–3.7 in $V$–$K$ color).[5] These ranges are based on work done by the RECONS group and Eric Mamajek's spectral notes.[6]

In the right panel of Figure 1 it is apparent that stars of the $\beta$ Pic moving group, the youngest sample examined here with ages $\sim$20 Myr (see Table 1), lie clearly above the main sequence. As evident in their spectra (described in Section 5), these pre-main-sequence stars exhibit significant chromospheric activity. They are larger and brighter because they are still settling onto the main sequence, as their outer layers are contracting and their internal temperatures are increasing. With ages of $\sim$40 Myr, K-dwarf members of the Tuc-Hor group are also noticeably above the main sequence but are elevated less than $\beta$ Pic K dwarfs. By ages of $\sim$145 Myr, members of the AB Dor moving group are found at positions indistinguishable from the main sequence. The Hyades cluster, field stars, and RV variable (RVV) stars all lie firmly on the main sequence, with the exception of the dwarf BD +05 2529, which has

---
[5] GAIA EDR3 Documentation: Relationships with other photometric systems
[6] http://www.pas.rochester.edu/~emamajek

**Table 1**
Moving Groups, Associations, and Clusters Providing K Dwarfs for the Benchmark Sample

| Group Name | R.A. (J2000) | Decl. (J2000) | Distance (pc)[a] | Age | Members[b,c] | K Dwarfs | Observed |
|---|---|---|---|---|---|---|---|
| $\beta$ Pictoris moving group | ~14 30 | ~−42 00 | ~30 | $24 \pm 3$ Myr[d] | 97 | 19 | 11 |
| Tuc-Hor association | ~02 36 | ~−52 03 | ~40 | $45 \pm 4$ Myr[d] | 176 | 18 | 10 |
| AB Doradus moving group | ~05 28 | ~−65 26 | ~33 | $145 \pm^{50}_{19}$ Myr[d] | 84 | 24 | 8 |
| Hyades cluster | ~04 26 | ~+15 52 | ~42 | $750 \pm 100$ Myr[d] | 177 | 47 | 10 |
| Field K dwarfs | Various | Various | <25 | 0.3–5.7 Gyr | ... | 10 | 5 |
| $o^2$ Eri | 04 15 16.3 | −07 39 10 | 5 | 4.3 Gyr[e] | ... | ... | ... |
| HD 50281 | 06 52 18.1 | −05 10 25 | 9 | 1.9 Gyr[f] | ... | ... | ... |
| 20 Crt | 11 34 29.5 | −32 49 53 | 10 | 4.6 Gyr[e] | ... | ... | ... |
| PX Vir | 13 03 49.7 | −05 09 43 | 22 | 0.3 Gyr[g] | ... | ... | ... |
| $\epsilon$ Ind | 22 03 21.7 | −56 47 10 | 4 | 3.7–5.7 Gyr[h] | ... | ... | ... |

**Notes.**
[a] Distance from Gaia EDR3.
[b] Membership list for $\beta$ Pic, Tuc-Hor, and AB Dor: Bell et al. (2015).
[c] Membership list for Hyades: Gagne et al. (2018).
[d] Gagne et al. (2018).
[e] Mamajek & Hillenbrand (2008).
[f] Luck (2017).
[g] Stanford-Moore et al. (2020).
[h] Feng et al. (2019).

recently been found to be a spectroscopic binary (Sperauskas et al. 2019) and is confirmed via our CHIRON data. Thus, only for ages less than ~50 Myr can young K dwarfs be identified via their positions on the H-R diagram, at least relative to the ensemble of mixed stars in the solar neighborhood.

### 2.1. Primary Sample of Field K Dwarfs

Of the ~5000 K-dwarf systems[7] in the full 50 pc sample, the first author's Ph.D. work focuses on the ~1200 systems within 40 pc and lying in the equatorial band of the sky from decl. +30° to −30°. These stars have been selected using Gaia parallax measurements (Gaia Collaboration et al. 2016, 2018; Lindegren et al. 2018) and continue to be vetted to provide, ultimately, a volume-limited and volume-complete sample. Sample construction continues because parallaxes may change slightly in future Gaia Data Releases and new K dwarfs may be added because solutions are not yet available. Some stars will be removed because the goal is to include systems with K-dwarf primaries, i.e., if white dwarfs are found, the system will be dropped from the primary sample because the white dwarf progenitor was originally more massive. We note that systems with white dwarf primaries are useful for age determinations, so these systems will continue to be considered for age calibration work but not included in the statistics for K-dwarf samples and their companions. There are more than 1200 K dwarfs in the equatorial 40 pc sample, of which more than 95% have been observed at least once between 2017 June and 2021 December.

Among the observed stars is a subset of 300 stars with repeated CHIRON observations targeted in an RV survey for companions (Paredes et al. 2021). We selected seven stars that exhibited RV variations of 50 m s$^{-1}$ or more that did not appear to be due to companions in the available CHIRON data as of 2018 December; instead, the variable velocities may be indicative of activity related to youth. Here we investigate these seven because they are promising candidates to be nearby young stars, and we evaluate them in comparison with the features examined in the spectra of the young stars in the benchmark sample.

### 2.2. Benchmark Sample of K Dwarfs with Age Estimates

A supplementary benchmark sample of ~100 K dwarfs includes those with age estimates. These stars have been taken from moving groups, associations, or clusters, plus a handful of field stars within 25 pc that have ages determined via isochrone fitting. Table 1 lists the various subsets used to construct the benchmark sample and the estimated ages of each group. The four associations utilized here are the $\beta$ Pictoris moving group ($\beta$ Pic, age $24 \pm 3$ Myr), the Tucana-Horologium association (Tuc-Hor, $45 \pm 4$ Myr), the AB Doradus moving group (AB Dor, $145 \pm^{50}_{19}$ Myr), and the Hyades cluster ($750 \pm 100$ Myr; Bell et al. 2015; Brandt & Huang 2015). The small set of field K dwarfs within 25 pc having age estimates made via model isochrone fits have ages of 0.3–5.7 Gyr.

K-dwarf members of the $\beta$ Pic, Tuc-Hor, and AB Dor groups were identified using the Bell et al. (2015) bona fide membership list and confirmed through the Gagne et al. (2018) web-based BANYAN $\Sigma$ code,[8] in combination with Gaia DR2 parallax, proper-motion, and RV data. This yielded a larger sample of K dwarfs than Gagne et al. (2018), which listed bona fide members before the Gaia DR2 release. Hyades membership was determined using K dwarfs in the Gagne et al. (2018) bona fide list with all 47 stars checked using BANYAN $\Sigma$ and updated Gaia DR2 data.

To date, 44 members of the benchmark sample have been observed using CHIRON, with 35 of these spectra having sufficient signal-to-noise ratios (S/Ns) for detailed spectral analysis. The discrepancy between the number of observed and the number of measured spectra mainly stems from the poor S/N achieved for fainter ($V > 12$), usually cooler stars. The EWs acquired from the set of 35 stars proved to be sufficient for identifying spectral age and activity trends.

---
[7] Defined to have a K-dwarf primary, plus any additional lower-mass stellar, brown dwarf, or planetary companions.

[8] BANYAN $\Sigma$ - http://www.exoplanetes.umontreal.ca/banyan/.

## 3. Observations and Data Reduction

All spectra used in this study were acquired between 2017 June and 2020 March using the CHIRON echelle spectrograph at the SMARTS 1.5 m telescope described in Tokovinin et al. (2013) and Paredes et al. (2021). We utilized CHIRON's slicer mode to attain spectra with resolution $R = 80,000$, with each spectrum split into 59 orders and covering a wavelength range of 4150–8800 Å. Each observation consisted of a single exposure of 900 s (stars with $V < 10.5$) or 1200 s ($V > 10.5$). This was done to ensure that we attained an S/N greater than 30 for all spectra. Each observation was followed by a single ThAr lamp exposure of 0.4 s that was used for wavelength calibration. Included with each night's observations were two sets of calibration frames integral to the data reduction process; one set is taken each afternoon before nighttime operations begin, and another after all observations have been completed for the night.

All CHIRON data are reduced using a customized data reduction pipeline described in Tokovinin et al. (2013), with additional details specific to our program in Paredes et al. (2021). The pipeline is currently run by members of the RECONS team, with CHIRON spectra reduced and distributed to dozens of research teams for over 1000 nights by the end of 2021. Briefly, each spectrum is bias-corrected and flat-fielded using quartz lamp calibrations to remove electronic readout noise and to correct for individual pixel sensitivities. After removing cosmic rays from the spectrum, profile order extraction is performed using an extraction algorithm based on the REDUCE package by Piskunov & Valenti (2002). Finally, each spectrum with extracted orders is matched with its closest ThAr calibration frame to obtain the sampled wavelength solution.

Once the basic pipeline reductions are done, the S/N per pixel is calculated between 6717 and 6720 Å using the method described in Tokovinin et al. (2013). Spectra with an S/N less than 30 are omitted from the present analysis because their EW measurements are often unreliable for the spectral features of interest. Each order is then trimmed at the edges to eliminate poor signal portions of the orders, and the spectra are normalized and flattened using a MATLAB script. Spectra are then shifted twice: (1) first by applying a barycentric velocity correction, and (2) then shifted to zero velocity relative to the Sun. Spectral analysis of specific lines was then carried out using the methods listed below.

## 4. Spectral Analysis

### 4.1. Line Selection

Four spectral features have been selected to create a rubric to evaluate ages and activity for K stars—H$\alpha$ at 6563 Å, the Na I doublet at 5890 and 5896 Å, Li I at 6708 Å, and Ca II at 8452 Å. Plots of the spectral regions containing the four selected lines for all 42 stars are shown in Figures 2 and 3.

The most direct information about stellar age for K dwarfs can be gleaned from a Li I resonance line at 6707.8 Å. Depletion of lithium in late-type dwarfs has been well documented in previous studies, such as Soderblom & Jones (1993), White et al. (2007), López-Santiago et al. (2010), and Binks & Jeffries (2014). It is proposed that lithium is destroyed as a young K star settles onto the main sequence through a process similar to the proton–proton chain reaction, with the end product being two helium atoms and the release of energy (Soderblom 2010). The decrease in the EW of the Li I line has been associated with increased age and is used in our study of K dwarfs as the most direct age marker (Soderblom et al. 2014). However, using the Li I feature is somewhat limiting because the majority of lithium is depleted within the first 200 Myr for dwarfs of type late G through early M (Soderblom et al. 2014). The trend in the Li I $\lambda$6707.8 line strength is also temperature dependent, fading faster for cool, late-type dwarfs (K8V–M9V), as shown by Riedel et al. (2017), who find an absence of Li I features in M-dwarf members of moving associations with age estimates of only 50 Myr.

To enhance our efforts to estimate ages, we also consider spectral lines resulting from stellar activity. Increased activity in the chromospheres of late-type stars has been linked with age since the publication of Skumanich (1972) five decades ago, so a comprehensive literature search was done to find spectral lines that might be used as activity, and presumably age, markers for K dwarfs. Candidate lines need to be located within CHIRON's spectral range of 4150–8800 Å, and therefore the popular Ca II H and K lines at 3968 and 3934 Å that trace chromospheric activity are excluded. For this study, we have identified the H$\alpha$ line at 6563 Å and one line of the Ca II infrared triplet at 8542 Å as activity tracers; both lines exhibit core emission or filled-in profiles when a K dwarf's chromosphere is active (Montes & Martin 1998). The other two Ca II infrared triplet lines at 8498 and 8662 Å are omitted because the orders produced by CHIRON's slicer mode are truncated at longer wavelengths and miss both lines.

Surface gravity diagnostic lines have also been proposed as age markers for K dwarfs (Soderblom 2010). The idea is that younger stars with ages <100 Myr are still contracting and have bloated atmospheres compared to their older counterparts already on the main sequence. Thus, the younger star's larger radius at the same mass results in a lower surface gravity, and this can be revealed via relatively narrower spectral lines. In effect, increased opacity in a fully contracted main-sequence star leads to more atomic collisions and interactions in its atmosphere, resulting in a wider absorption feature with broader wings—this process is called pressure broadening. The Na I doublet lines at 5889.95 and 5895.92 Å are very sensitive to pressure broadening. An increase in the EW of the Na I doublet feature (EW[Na I D]) is therefore theorized to accompany an increase in age (Soderblom 2010).

### 4.2. Equivalent Width Measurements

To carry out the spectroscopic analysis of age and activity for the sample stars, we measured the EWs of both Na I lines at 5889.95 and 5895.92 Å, the H$\alpha$ line at 6563 Å, the unresolved Li doublet at 6707.8 Å, and the Ca II infrared triplet line at 8542 Å, using the SPLAT-VO software, which is distributed by the Starlink Project (Škoda et al. 2014). We compared SPLAT-VO to other comparable methods for measuring EWs, including the SPLOT package in IRAF, specutils using Python, and VOSpec from the European Space Agency. All methods resulted in similar EWs for a test sample of K dwarfs, and we decided to use SPLAT-VO owing to its user-friendly interface.

All 42 K dwarfs were analyzed, including 35 from the benchmark sample and the 7 RVV stars. A 20 Å window was created for the Na I doublet feature, and a 10 Å window was created around the centers of the Li, H$\alpha$, and Ca II lines, with pseudocontinuum fits made across each window using the normalized data before carrying out the EW measurements. SPLAT-VO uses the ABLINE technique to fit a Gaussian,

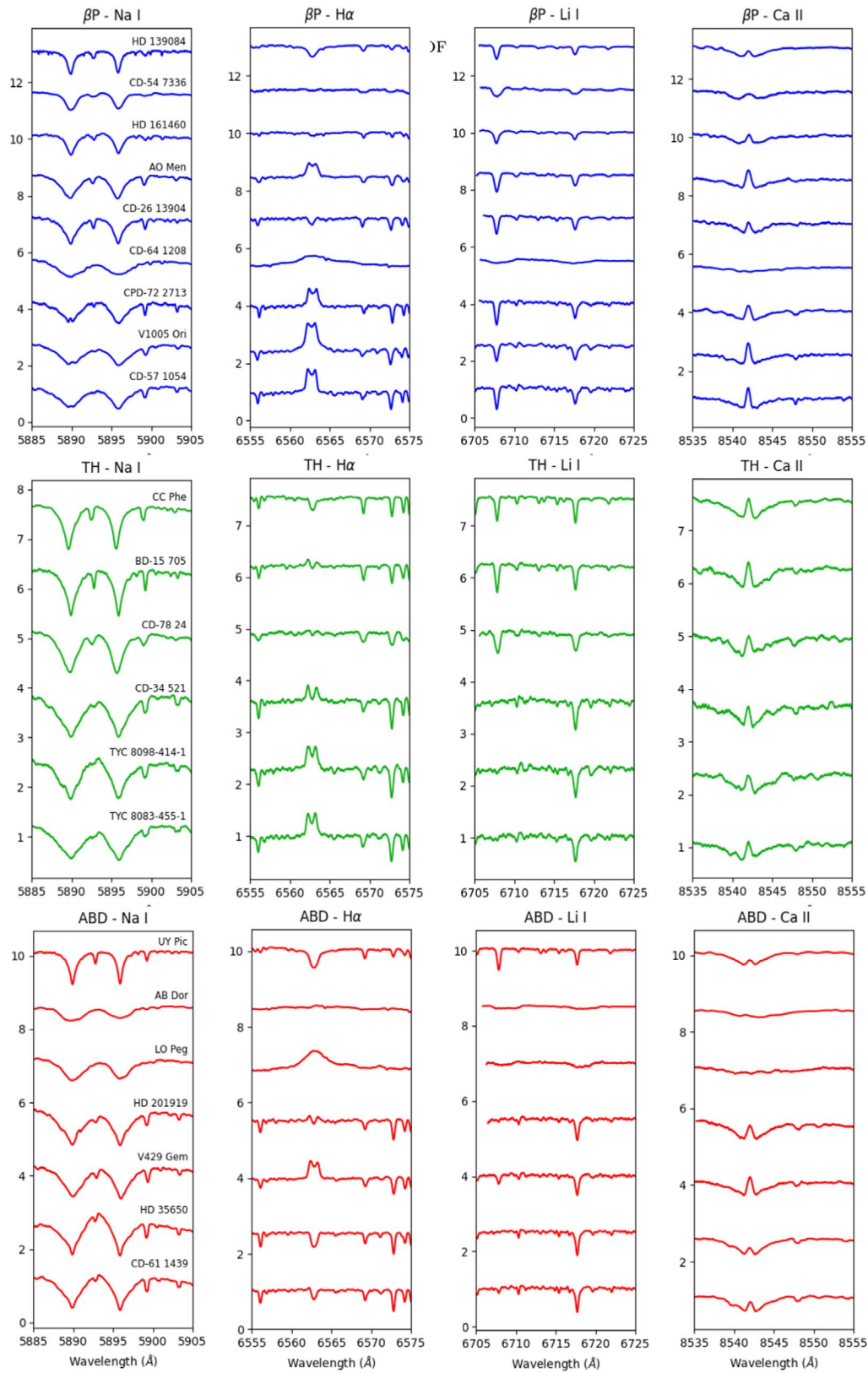

**Figure 2.** Top: compilation of nine K-dwarf spectra from the $\beta$ Pic moving group, focusing on the Na I doublet at 5889.95 and 5895.92 Å (first column), the H$\alpha$ line at 6562.8 Å (second column), the Li I absorption line at 6707.8 Å (third column), and one of the Ca II infrared triplet lines at 8542 Å (fourth column). Each plot spans 20 Å. Middle: compilation of six K-dwarf spectra from the Tuc-Hor association, focusing on the same features. Bottom: compilation of seven K-dwarf spectra from the AB Dor moving group, focusing on the same features.

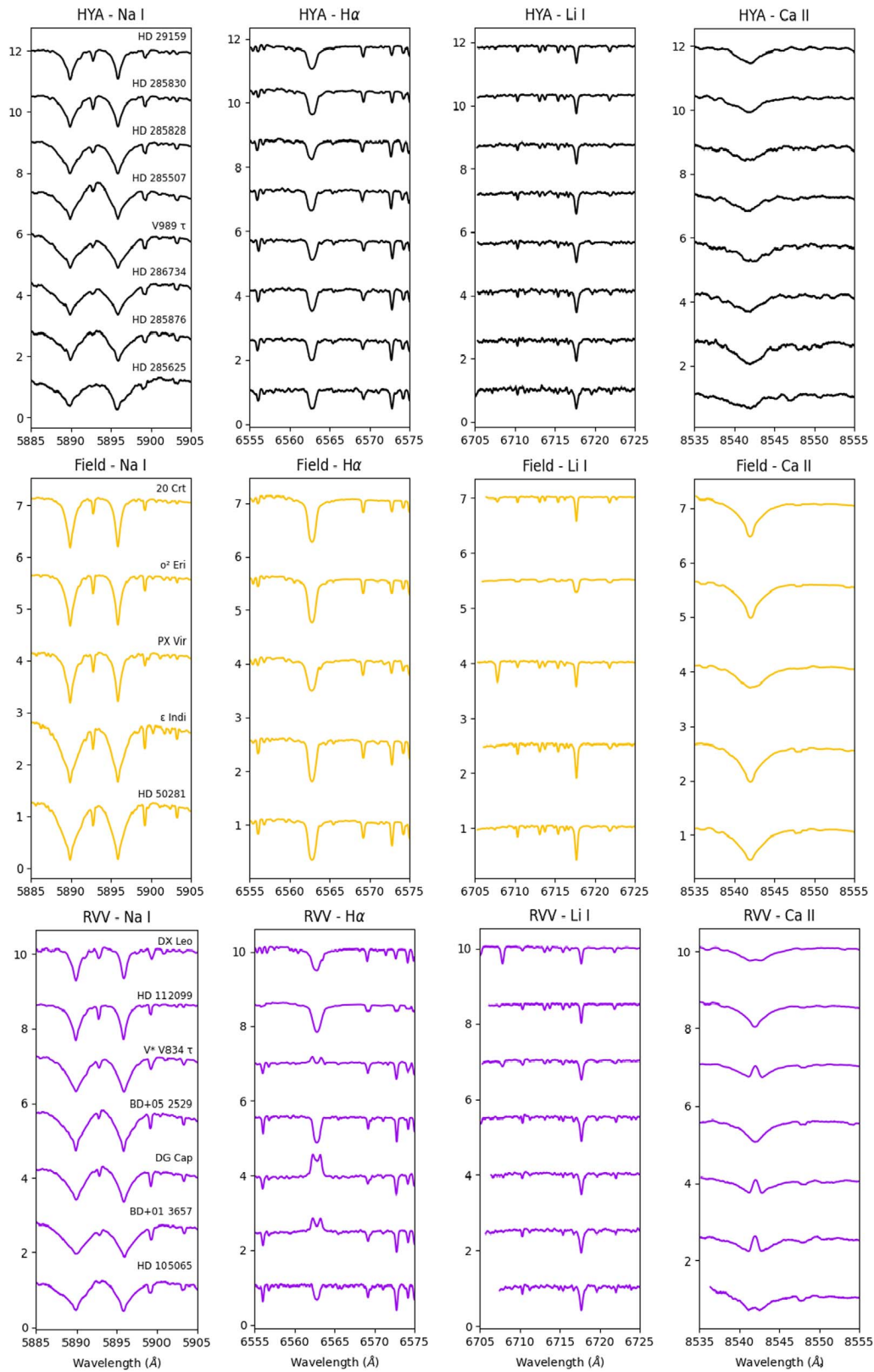

**Figure 3.** Top: compilation of eight K-dwarf spectra from the Hyades cluster, focusing on the Na I doublet at 5889.95 and 5895.92 Å (first column), the Hα line at 6562.8 Å (second column), the Li I absorption line at 6707.8 Å (third column), and one of the Ca II infrared triplet lines at 8542 Å (fourth column). Each plot spans 20 Å. Middle: compilation of five field K-dwarf spectra with known ages, focusing on the same features. Bottom: compilation of seven field K-dwarf spectra for stars that are RVVs, focusing on the same features.

Table 2
Spectroscopic Results and Derived Stellar Properties of the 42 Benchmark Sample Stars

| Name | Group | $T_{eff}$ (K) | [Fe/H] (dex) | log $g$ (dex) | $v \sin i$ (km s$^{-1}$) | EW[Na I D] (Å) | EW[H$\alpha$] (Å) | EW[Ca II] (Å) | EW[Li I] (Å) | S/N at Li |
|---|---|---|---|---|---|---|---|---|---|---|
| V1005 Ori | $\beta$P | 3944 ± 6 | 0.02 ± 0.13 | 4.68 ± 0.14 | 12.3 | 5.52 | −2.15 | 1.19 | 0.31 | 41 |
| CD −57 1054 | $\beta$P | 3937 ± 10 | −0.01 ± 0.12 | 4.68 ± 0.13 | <10 | 5.46 | −1.29 | 1.22 | 0.38 | 43 |
| AO Men | $\beta$P | 4641 ± 46 | −0.04 ± 0.04 | 4.56 ± 0.17 | 17.2 | 3.76 | −0.95 | 1.10 | 0.41 | 48 |
| HD 139084 | $\beta$P | 5333 ± 106 | −0.03 ± 0.16 | 4.52 ± 0.17 | 16.6 | 1.81 | 0.81 | 1.20 | 0.28 | 95 |
| CD −54 7336 | $\beta$P | 5226 ± 84 | 0.08 ± 0.09 | 4.42 ± 0.29 | 35.7 | 2.17 | −0.04 | 1.03 | 0.30 | 46 |
| HD 161460 | $\beta$P | 5293 ± 134 | −0.14 ± 0.19 | 4.55 ± 0.15 | 19.9 | 2.13 | 0.15 | 0.76 | 0.32 | 55 |
| CD −64 1208 | $\beta$P | NM | NM | NM | >50 | 4.63 | −2.80 | 0.66 | 0.32 | 68 |
| CD −26 13904 | $\beta$P | 4724 ± 25 | −0.10 ± 0.07 | 4.54 ± 0.16 | 12.7 | 3.80 | 0.16 | 0.79 | 0.31 | 39 |
| CPD −72 2713 | $\beta$P | 4085 ± 34 | −0.05 ± 0.14 | 4.68 ± 0.13 | <10 | 5.39 | −1.06 | 1.33 | 0.36 | 31 |
| CD −78 24 | TH | 4724 ± 73 | −0.18 ± 0.08 | 4.54 ± 0.19 | 18.6 | 3.54 | −0.04 | 0.80 | 0.27 | 28 |
| CD −34 52 1 | TH | 4001 ± 24 | −0.11 ± 0.12 | 4.70 ± 0.13 | <10 | 5.47 | −0.41 | 1.24 | NM | 25 |
| CC Phe | TH | 4969 ± 25 | −0.04 ± 0.02 | 4.54 ± 0.14 | <10 | 2.91 | 0.23 | 1.33 | 0.21 | 51 |
| BD −15 705 | TH | 4895 ± 32 | −0.06 ± 0.04 | 4.51 ± 0.15 | <10 | 2.90 | −0.17 | 1.41 | 0.23 | 36 |
| TYC 8083-455-1 | TH | 4024 ± 53 | −0.17 ± 0.15 | 4.69 ± 0.13 | <10 | 5.06 | −0.66 | 1.47 | 0.02 | 23 |
| TYC 8098-414-1 | TH | 4016 ± 34 | −0.07 ± 0.13 | 4.69 ± 0.13 | 10.3 | 6.15 | −0.74 | 1.51 | 0.02 | 21 |
| HD 35650 | ABD | 4191 ± 18 | −0.15 ± 0.07 | 4.67 ± 0.12 | 9.6 | 5.67 | 0.30 | 1.81 | 0.01 | 77 |
| AB Dor | ABD | NM | NM | NM | >50 | 2.27 | −0.77 | 1.18 | 0.27 | 129 |
| UY Pic | ABD | 5248 ± 58 | −0.09 ± 0.07 | 4.53 ± 0.15 | 12.5 | 2.01 | 0.90 | 1.40 | 0.26 | 77 |
| CD −61 1439 | ABD | 4160 ± 33 | −0.14 ± 0.09 | 4.68 ± 0.12 | 9.3 | 5.82 | 0.16 | 1.57 | 0.04 | 59 |
| V429 Gem | ABD | 4314 ± 25 | −0.23 ± 0.06 | 4.66 ± 0.13 | 12.5 | 4.65 | −0.65 | 1.39 | 0.11 | 39 |
| HD 201919 | ABD | 4343 ± 31 | −0.06 ± 0.18 | 4.64 ± 0.16 | <10 | 5.20 | −0.08 | 1.10 | 0.03 | 34 |
| LO Peg | ABD | NM | NM | NM | >50 | 3.70 | −2.90 | 0.89 | 0.10 | 42 |
| HD 285507 | HYA | 4479 ± 29 | 0.08 ± 0.02 | 4.60 ± 0.15 | <10 | 5.89 | 0.71 | 1.51 | NM | 30 |
| HD 285625 | HYA | 4068 ± 23 | 0.12 ± 0.51 | 4.67 ± 0.14 | <10 | 6.17 | 0.55 | 1.34 | NM | 21 |
| V989 $\tau$ | HYA | 4381 ± 11 | 0.07 ± 0.04 | 4.64 ± 0.12 | <10 | 6.01 | 0.70 | 1.39 | NM | 31 |
| HD 286734 | HYA | 4257 ± 17 | −0.04 ± 0.06 | 4.66 ± 0.13 | <10 | 6.46 | 0.64 | 1.77 | NM | 27 |
| HD 285828 | HYA | 4631 ± 33 | 0.06 ± 0.06 | 4.57 ± 0.16 | <10 | 5.03 | 0.65 | 1.08 | NM | 31 |
| HD 285830 | HYA | 5019 ± 29 | 0.22 ± 0.08 | 4.51 ± 0.13 | <10 | 3.33 | 0.91 | 1.90 | NM | 43 |
| HD 285876 | HYA | 4258 ± 9 | −0.03 ± 0.05 | 4.66 ± 0.13 | <10 | 6.29 | 0.64 | 2.72 | NM | 23 |
| HD 29159 | HYA | 5183 ± 29 | 0.09 ± 0.07 | 4.51 ± 0.15 | <10 | 2.61 | 0.82 | 1.26 | NM | 43 |
| $o^2$ Eri | Field | 5109 ± 32 | −0.43 ± 0.02 | 4.49 ± 0.17 | <10 | 2.39 | 1.22 | 1.73 | NM | 190 |
| HD 50281 | Field | 4710 ± 19 | 0.03 ± 0.02 | 4.54 ± 0.13 | <10 | 4.73 | 0.82 | 2.13 | NM | 178 |
| 20 Crt | Field | 5220 ± 46 | −0.47 ± 0.05 | 4.54 ± 0.12 | <10 | 2.19 | 1.31 | 1.70 | NM | 72 |
| PX Vir | Field | 5195 ± 52 | −0.12 ± 0.14 | 4.56 ± 0.14 | <10 | 2.35 | 0.87 | 1.63 | 0.15 | 100 |
| eps Ind | Field | 4617 ± 29 | −0.09 ± 0.03 | 4.58 ± 0.13 | <10 | 4.71 | 0.85 | 1.92 | NM | 116 |
| V* V834 $\tau$ ② | RVV | 4562 ± 18 | −0.19 ± 0.07 | 4.59 ± 0.01 | 12.6 | 4.34 | −0.11 | 1.45 | 0.06 | 93 |
| DX Leo ① | RVV | 5283 ± 54 | −0.09 ± 0.06 | 4.52 ± 0.02 | 10.4 | 2.06 | 0.77 | 1.55 | 0.18 | 113 |
| BD +05 2529 ⑦ | RVV | 4226 ± 35 | −0.28 ± 0.06 | 4.68 ± 0.01 | <10 | 6.04 | 0.64 | 1.84 | 0.01 | 54 |
| HD 105065 ⑤ | RVV | 4028 ± 32 | −0.16 ± 0.20 | 4.68 ± 0.02 | <10 | 5.76 | 0.05 | 1.09 | 0.01 | 39 |
| HD 112099 ⑥ | RVV | 5083 ± 49 | −0.12 ± 0.05 | 4.50 ± 0.15 | <10 | 2.61 | 0.94 | 1.63 | 0.01 | 92 |
| BD +01 3657 ④ | RVV | 4186 ± 10 | −0.19 ± 0.04 | 4.68 ± 0.01 | 10.6 | 5.42 | −0.55 | −0.11 | 0.01 | 41 |
| DG Cap ③ | RVV | 4092 ± 49 | −0.45 ± 0.06 | 4.70 ± 0.01 | 10.5 | 4.75 | −0.85 | 1.41 | 0.02 | 49 |

**Note.** NM: not measured. EW[Na I D]: EW of the Na I doublet at 5889.95 and 5895.92 Å. EW[H$\alpha$]: EW of the H$\alpha$ line at 6563 Å. RVV stars are numbered to match the numbers in Figure 9 below.

Lorentzian, or Voigt profile (depending on user preference) to the absorption feature and to measure the EW of the line (Škoda et al. 2014).[9] Gaussians were used for both lines of the Na I doublet, which were measured individually before summing their EWs to produce the EW[Na I D] values reported here. Ca II lines were also measured using a Gaussian because of the broadened wings observed in the line profiles, while Voigt profiles were preferred for the Li and H$\alpha$ lines. Results are given in Table 2. Errors have been estimated for all EWs as discussed in Section 5.3.

---
[9] The SPLAT-VO User Manual can be found at http://star-www.dur.ac.uk/~pdraper/splat/splat-vo/sun243.pdf.

### 4.3. Stellar Parameters

Empirical SpecMatch (ESM) from Yee et al. (2017) was used to derive stellar properties for all 42 K dwarfs in this study, with results given in Table 2. Using a library of 404 late-type stars with high-resolution spectra and reliable stellar properties derived from interferometry, asteroseismology, LTE spectral synthesis, and spectrophotometry, ESM can quickly and accurately estimate values through spectral comparison (Yee et al. 2017). When a new target spectrum is introduced, ESM's Python code produces a linear combination of the best-matching library spectra and uses a combination of their stellar values to produce an estimate. Yee et al. (2017) report

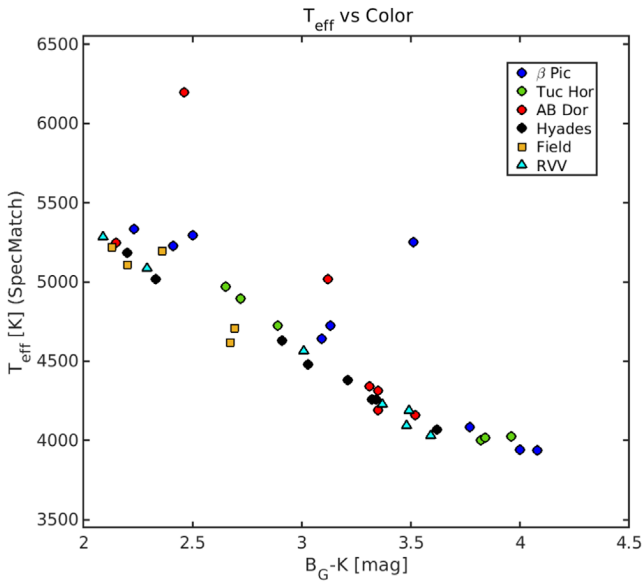

**Figure 4.** Effective temperature derived from Empirical SpecMatch plotted against $B_G - K$ color for all 42 K dwarfs in our sample.

uncertainties of 70 K in effective temperature and 0.12 dex in [Fe/H] for mid- to late K dwarfs.

ESM focuses on the optical wavelength range from 5100 to 5800 Å, which is well suited to the CHIRON spectra acquired for this study. This range includes the Mg $b$ triplet lines located at 5140–5190 Å that are excellent for stellar spectral analysis (Yee et al. 2017; Morris et al. 2019) and avoids the telluric lines found from 6270 to 6310 Å.

Our analysis focused on five regions from 5140 to 5800 Å, chosen to avoid areas of increased noise in the CHIRON spectra. The five regions span wavelengths 5340–5400 Å, 5440–5500 Å, 5540–5600 Å, 5640–5700 Å, and 5740–5800 Å. We derived effective temperature ($T_{\rm eff}$), metallicity ([Fe/H]), and log $g$ values in each of the five regions and then averaged the values to arrive at a final estimate for each star. Errors on each value represent the standard deviations of the five values. Systematic error analyses for both $T_{\rm eff}$ and [Fe/H] are discussed in Section 5.4. Rotational velocity ($v \sin i$) values were determined by comparing spectra to spectral standards with well-known rotational velocities in the ESM library (Yee et al. 2017). The weighted average of five $v \sin i$ standards was used to estimate each star's $v \sin i$; these values are reported in Table 2. The Empirical SpecMatch library limited the range of $v \sin i$ values with explicit estimates, with a lack of fast rotors with $v \sin i$ greater than 50 km s$^{-1}$.

Figure 4 shows a comparison between the ESM-derived effective temperatures and the $B_G - K$ colors of the stars. There is a clear trend for most of the 42 stars in our sample. The three most extreme outliers are CD –64 1208, AB Dor, and LO Peg —these are all fast rotators with extremely wide absorption-line profiles that compromise $T_{\rm eff}$ measurements.

### 4.4. Gamma Velocity and UVW Space Motion Analysis

We measure the gamma ($\gamma$) velocities (i.e., systemic RVs) from the spectra of the 42 K dwarfs using our RV pipeline, described in more detail in Paredes et al. (2021). The order of operations for this effort is the same as outlined above through the barycenter motion correction, which is assigned using the Barycorr package provided by Wright & Eastman (2014) that requires the geographical coordinates of CHIRON, the time stamp of the observation (middle of the exposure used), and the star's astrometric location on the sky. Next, a template spectrum is chosen for wavelength grid matching, and the two spectra are cross-correlated. The $\gamma$ velocity is derived from a standard set of 14 of the 59 echelle orders produced in slicer mode: orders 10, 12, 13, 16, 17, 18, 20, 21, 22, 23, 24, 27, 30, and 35. Uncertainties for the orders are based on the shapes of the cross-correlation functions (CCFs) and the S/N in each, and the final $\gamma$ velocity value and error are computed using the weighted average and standard deviation of the 14 order values. For most stars, propagation of 1$\sigma$ errors in distance, proper motions, and $\gamma$ velocities in combination does not result in changes in UVW velocities by more than 1 km s$^{-1}$ in any individual velocity value, a result of the very high quality astrometric and RV data. For a few very fast rotators with more poorly determined $\gamma$ values, some UVW errors are 2–3 km s$^{-1}$.

In order to confirm group membership, we determine the UVW Galactic space motions of each system. These are right-handed Cartesian velocities, where the $U$-axis is positive toward the Galactic center, the $V$-axis is positive in the direction of Galactic rotation, and the $W$-axis is positive toward the north Galactic pole. Here we utilize the gal_uvw routine from the Python AstroLib set of libraries, where we input each object's celestial coordinates, proper motions, and distance via parallax, all taken from Gaia EDR3, plus our derived $\gamma$ velocity (all values given in Table 3). Note that we have not removed the solar motion in this computation, so the values are not corrected for the local standard of rest (LSR). Results for $\gamma$ velocities and UVW space motions are given in Table 3. All of these motions are useful in confirming or refuting individual stars' memberships to moving groups or associations because we expect to see similar UVW values and $\gamma$ velocities for members of the same kinematic group.

### 5. Results

Figures 2 and 3 show a compilation of K-dwarf spectra from all five benchmark groups (see Table 1), presented in order of age, with the seven RVV stars in the final set of panels. Within each group, the stellar spectra are arranged based on $B_G$–$K$ color increasing from top to bottom, effectively hotter to cooler $T_{\rm eff}$. It is clear at first glance that several stars exhibit very broad features indicative of fast rotation, e.g., CD −64 1208, AB Dor, and LO Peg. The specific spectral features highlighted are the Na I doublet highlighted in the first column, the H$\alpha$ line in the second column, the Li I region in the third column, and the Ca II feature in the fourth column. There are subtle trends in the Na I doublet feature that are not evident to the eye but will be discussed below in Section 5.1. In contrast, an obvious trend is observed for the H$\alpha$ line, which exhibits core emission or at least partially filled-in line profiles for all of the K dwarfs in $\beta$ Pic, Tuc-Hor, and AB Dor, but comparatively deep absorption lines typical of older K dwarfs for the Hyades cluster members and the five nearby stars with ages from isochrones. We observe strong Li I lines in all members of $\beta$ Pic and the hotter members of Tuc-Hor, while the line is generally weak or nonexistent in AB Dor and Hyades members. One of the field K dwarfs, PX Vir, shows a strong Li I absorption feature, which is likely why it has been classified in the past as an AB Dor member (Bell et al. 2015). Among the seven RVV stars, DX Leo and V$^*$ V834 $\tau$ show Li I absorption features. Finally, the Ca II features are generally correlated with H$\alpha$—stars with

**Table 3**
Dynamical and Color Information for Sample Stars

| Name | [a]Group | R.A. (J2000.0) | Decl. (J2000.0) | $\mu_{R.A}$ (mas yr$^{-1}$) | $\mu_{Decl}$ (mas yr$^{-1}$) | $\pi$ (mas) | $\sigma_\pi$ (mas) | $B_{Gaia}$ (mag) | $B_G - K_S$ (mag) | $\gamma_{REC}$ (km s$^{-1}$) | $\sigma_{\gamma_{REC}}$ (km s$^{-1}$) | $\gamma_{Gaia}$ (km s$^{-1}$) | $U$ (km s$^{-1}$) | $V$ (km s$^{-1}$) | $W$ (km s$^{-1}$) |
| --- | --- | --- | --- | --- | --- | --- | --- | --- | --- | --- | --- | --- | --- | --- | --- |
| (1) | (2) | (3) | (4) | (5) | (6) | (7) | (8) | (9) | (10) | (11) | (12) | (13) | (14) | (15) | (16) |
| V1005 Ori | βP | 04 59 34.8 | +01 47 01 | 39.130 | −94.900 | 40.990 | 0.013 | 10.26 | 4.00 | 18.832 | 0.198 | 18.079 | −12.2 | −16.2 | −9.2 |
| CD −57 1054 | βP | 05 00 47.1 | −57 15 25 | 35.388 | 74.113 | 37.212 | 0.013 | 10.33 | 4.08 | 19.077 | 0.105 | 18.073 | −11.1 | −16.4 | −9.0 |
| AO Men | βP | 06 18 28.2 | −72 02 41 | −7.709 | 74.412 | 25.566 | 0.013 | 9.91 | 3.09 | 16.575 | 0.435 | 10.952 | −10.4 | −16.8 | −8.8 |
| HD 139084 | βP | 15 38 57.6 | −57 42 27 | −54.602 | −92.786 | 25.829 | 0.201 | 8.08 | 2.23 | −0.286 | 0.483 | 3.900 | −11.2 | −14.4 | −7.7 |
| CD −54 7336 | βP | 17 29 55.1 | −54 15 48 | −5.490 | −63.435 | 14.790 | 0.014 | 9.78 | 2.41 | 0.360 | 0.307 | 1.6[b] | −8.4 | −16.2 | −9.1 |
| HD 161460[c] | βP | 17 48 33.7 | −53 06 43 | −6.730 | −57.235 | 13.053 | 0.114 | 9.28 | 2.50 | 1.321 | 0.237 | 11.189 | −7.3 | −18.0 | −7.9 |
| CD −64 1208 | βP | 18 45 36.9 | −64 51 47 | 19.891 | −154.401 | 35.158 | 0.181 | 9.61 | 3.51 | −2.071 | 5.739 | 7.939 | −13.9 | −14.7 | −6.0 |
| CD −26 13904 | βP | 19 11 44.7 | −26 04 08 | 21.447 | −48.196 | 14.769 | 0.212 | 10.50 | 3.13 | −9.465 | 0.186 | −8.1[b] | −10.0 | −13.8 | −9.3 |
| CPD −72 2713 | βP | 22 42 48.9 | −71 42 21 | 94.854 | −52.384 | 27.231 | 0.011 | 10.67 | 3.77 | 8.006 | 0.081 | 7.019 | −10.3 | −15.8 | −8.0 |
| CD −78 24 | TH | 00 42 20.3 | −77 47 39 | 79.965 | −29.881 | 20.084 | 0.009 | 10.42 | 2.89 | 11.159 | 0.531 | 10.994 | −9.0 | −21.1 | −2.3 |
| CD −34 521 | TH | 01 22 04.4 | −33 37 03 | 109.930 | −57.349 | 25.904 | 0.016 | 11.27 | 3.82 | 4.719 | 0.074 | 3.985 | −9.9 | −20.9 | −1.1 |
| CC Phe | TH | 01 28 08.6 | −52 38 19 | 106.287 | −43.149 | 25.110 | 0.011 | 9.49 | 2.65 | 8.052 | 0.054 | 7.825 | −9.7 | −20.9 | −0.9 |
| BD −15 705 | TH | 04 02 16.4 | −15 21 29 | 66.103 | −26.834 | 18.282 | 0.013 | 10.30 | 2.72 | 15.502 | 0.109 | 14.553 | −11.6 | −21.1 | −1.5 |
| TYC 8083-455-1 | TH | 04 48 00.6 | −50 41 25 | 56.404 | 19.164 | 17.194 | 0.023 | 11.88 | 3.96 | 19.331 | 0.070 | 19.3[b] | −11.6 | −22.5 | −0.9 |
| TYC 8098-414-1 | TH | 05 33 25.5 | −51 17 13 | 42.970 | 26.107 | 18.560 | 0.012 | 12.00 | 3.84 | 20.448 | 0.128 | 19.472 | −10.5 | −21.7 | −1.5 |
| HD 35650 | ABD | 05 24 30.1 | −38 58 10 | 43.159 | −57.276 | 57.271 | 0.015 | 9.27 | 3.35 | 32.330 | 0.057 | 32.078 | −8.0 | −28.0 | −15.3 |
| AB Dor | ABD | 05 28 44.9 | −65 26 55 | 37.554 | 158.574 | 67.333 | 0.441 | 7.15 | 2.46 | 19.191 | 0.868 | 29.500 | −9.8 | −17.8 | −9.3 |
| UY Pic | ABD | 05 36 56.9 | −47 57 53 | 23.129 | −1.134 | 40.657 | 0.014 | 7.96 | 2.15 | 32.596 | 0.157 | 32.359 | −7.3 | −28.1 | −15.0 |
| CD −61 1439 | ABD | 06 39 50.0 | −61 28 41 | −26.991 | 75.011 | 45.033 | 0.012 | 10.02 | 3.52 | 31.617 | 0.050 | 31.576 | −7.7 | −28.3 | −14.5 |
| V429 Gem | ABD | 07 23 43.5 | +20 24 58 | −65.642 | −230.692 | 36.086 | 0.019 | 10.23 | 3.35 | 7.988 | 0.167 | 7.066 | −5.1 | −27.2 | −17.0 |
| HD 201919 | ABD | 21 13 05.2 | −17 29 12 | 79.282 | −146.185 | 26.021 | 0.016 | 10.89 | 3.31 | −7.394 | 0.108 | −8.502 | −6.9 | −27.4 | −13.3 |
| LO Peg | ABD | 21 31 01.7 | +23 20 07 | 134.654 | −144.889 | 41.291 | 0.017 | 9.50 | 3.12 | −41.571 | 1.664 | −23.355 | −11.2 | −45.5 | −7.0 |
| HD 285507 | HYA | 04 07 01.2 | +15 20 06 | 124.626 | −19.786 | 22.229 | 0.017 | 10.70 | 3.03 | 37.912 | 0.046 | 38.121 | −42.6 | −18.7 | −1.5 |
| HD 285625 | HYA | 04 15 10.4 | +14 23 54 | 114.581 | −18.388 | 20.978 | 0.018 | 11.75 | 3.62 | 38.267 | 0.051 | 37.622 | −42.1 | −19.4 | −1.0 |
| V989 τ | HYA | 04 23 25.2 | +15 45 47 | 125.173 | −26.474 | 24.134 | 0.025 | 10.70 | 3.21 | 38.408 | 0.046 | 38.423 | −41.6 | −19.3 | −0.6 |
| HD 286734 | HYA | 04 23 54.4 | +14 03 07 | 115.516 | −19.066 | 23.111 | 0.019 | 11.10 | 3.32 | 39.637 | 0.040 | 39.545 | −42.4 | −18.7 | −1.5 |
| HD 285828 | HYA | 04 27 25.3 | +14 15 38 | 102.458 | −19.956 | 20.422 | 0.192 | 10.62 | 2.91 | 44.302 | 0.042 | 39.248 | −46.4 | −19.6 | −3.0 |
| HD 285830 | HYA | 04 27 47.0 | +14 25 04 | 100.550 | −19.200 | 20.353 | 0.016 | 9.68 | 2.33 | 39.576 | 0.049 | 39.592 | −42.0 | −19.1 | −1.2 |
| HD 285876 | HYA | 04 31 52.4 | +15 29 58 | 105.365 | −24.241 | 21.872 | 0.016 | 11.24 | 3.34 | 40.315 | 0.044 | 40.443 | −42.5 | −19.1 | −1.2 |
| HD 29159 | HYA | 04 36 05.2 | +15 41 02 | 95.399 | −24.022 | 19.993 | 0.018 | 9.57 | 2.20 | 40.784 | 0.052 | 40.619 | −42.7 | −19.6 | −1.0 |
| $o^2$ Eri | Field | 04 15 16.3 | −07 39 10 | −2240.085 | −3421.809 | 199.608 | 0.121 | 4.61 | 2.20 | −42.325 | 0.038 | −42.621 | 96.8 | −12.4 | −41.3 |
| HD 50281 | Field | 06 52 18.1 | −05 10 25 | −543.690 | −3.515 | 114.355 | 0.042 | 6.80 | 2.69 | −7.057 | 0.038 | −7.204 | 0.0 | 12.8 | −19.9 |
| 20 Crt | Field | 11 34 29.5 | −32 49 53 | −670.230 | 822.399 | 104.613 | 0.028 | 6.15 | 2.13 | −21.912 | 0.040 | −21.974 | −47.5 | 19.5 | 12.3 |
| PX Vir | Field | 13 03 49.7 | −05 09 43 | −191.130 | −218.730 | 46.100 | 0.810 | 7.87 | 2.36 | 9.149 | 0.074 | −7.547 | −2.9 | −30.9 | −3.3 |
| ε Indi | Field | 22 03 21.7 | −56 47 10 | 3966.661 | −2536.192 | 274.843 | 0.096 | 4.91 | 2.67 | −40.096 | 0.035 | −40.504 | −80.8 | −40.9 | 2.3 |
| V* V834 τ | RVV | 04 41 18.9 | +20 54 05 | −234.261 | −254.314 | 75.687 | 0.024 | 8.16 | 3.01 | 0.995 | 0.170 | 0.546 | 5.0 | −3.5 | −20.8 |
| DX Leo | RVV | 09 32 43.8 | +26 59 19 | −147.260 | −246.593 | 55.329 | 0.021 | 7.21 | 2.09 | 8.309 | 0.087 | 7.936 | −10.1 | −23.3 | −5.6 |
| BD +05 2529 | RVV | 11 41 49.6 | +05 08 26 | 230.643 | −469.130 | 31.762 | 0.381 | 9.80 | 3.37 | 18.909 | 0.040 | 19.141 | 62.7 | −50.2 | −0.5 |
| HD 105065 | RVV | 12 05 50.7 | −18 52 31 | −15.018 | −317.780 | 43.365 | 0.092 | 10.21 | 3.59 | −53.945 | 0.051 | −29.157 | 1.2 | 17.8 | −61.7 |
| HD 112099 | RVV | 12 53 54.4 | +06 45 46 | −231.846 | 93.538 | 36.594 | 0.061 | 8.44 | 2.29 | −26.576 | 0.039 | −18.923 | −36.6 | 0.7 | −20.4 |
| BD +01 3657 | RVV | 18 22 17.2 | +01 42 25 | 84.197 | −19.688 | 37.864 | 0.017 | 10.35 | 3.49 | −16.972 | 0.088 | −16.802 | −14.6 | −5.8 | −12.6 |
| DG Cap | RVV | 20 41 42.2 | −22 19 20 | 656.071 | −538.708 | 41.326 | 0.042 | 10.08 | 3.48 | −46.001 | 0.085 | −62.133 | −68.1 | −66.6 | −50.3 |

**Notes.** Column (1): object identifier. Column (2): sample. Column (3): right ascension. Column (4): declination. Columns (5) and (6): proper motion. Column (7): trigonometric parallax. Column (8): error in trigonometric parallax. Column (9): Gaia EDR3 BP magnitude = $B_{Gaia}$. Column (10): long baseline color. Column (11): CHIRON $\gamma$ velocity. Column (12): error in CHIRON $\gamma$ velocity. Column (13): Gaia DR2 $\gamma$ velocity. Columns (14)–(16): U, V, and W space velocity components.

[a] Moving groups located in the second column are denoted by the following: βP = Beta Pictoris; TH = Tucana-Horologium; ABD = AB Doradus; HYA = Hyades; Field = K field star; RVV = RV variable star.
[b] Measurements obtained from Torres et al. (2006).
[c] Astrometry obtained from Gaia DR2.

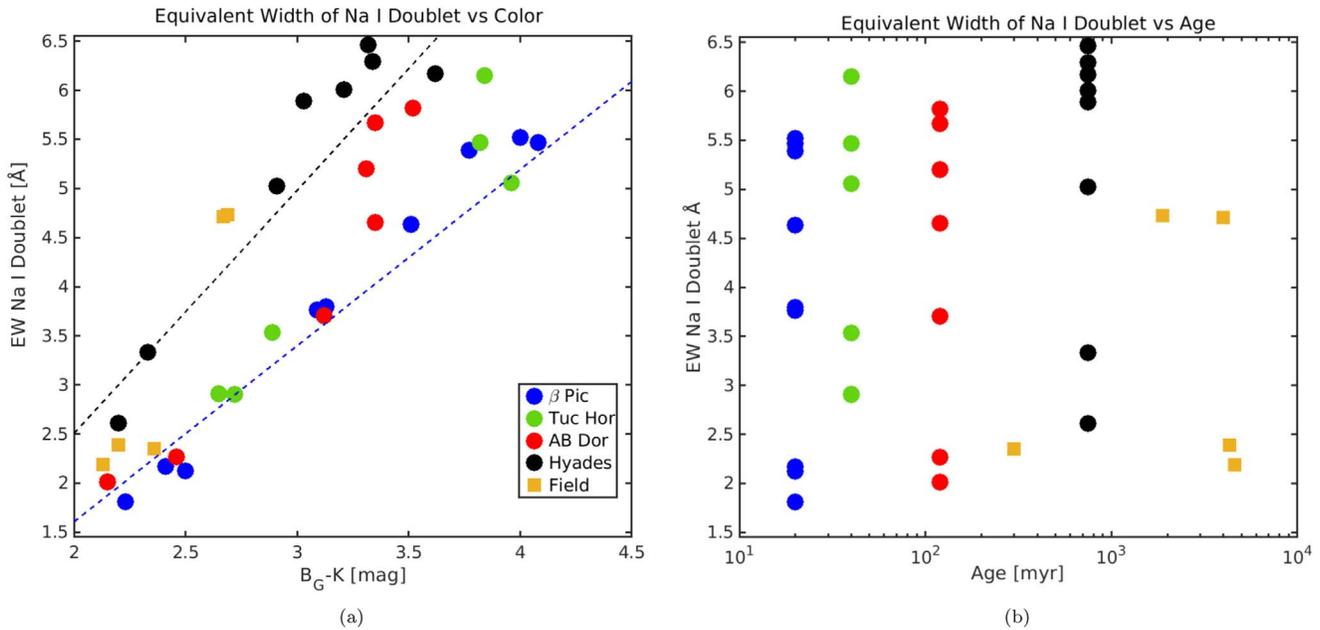

**Figure 5.** (a) Plot showing the EW[Na I D] vs. $B_G$–$K$ color for the five groups of stars used in the benchmark sample. Cluster trends are indicated by lines of the same color. (b) Plot showing the EW[Na I D] vs. estimated age for all five benchmark age groups.

strong H$\alpha$ emission typically show strong re-emission features in the cores of their Ca II lines.

### 5.1. Benchmark Sample Results

Table 2 presents spectral parameter results and EWs for the key features for the entire sample separated into the various groups, within which stars are listed in order of R.A. The extracted EWs for Na I, H$\alpha$, the Ca II triplet line, and Li I are given, as well as the S/N of the spectra near the Li I feature (computed as described above). In combination, the EWs of these spectral features provide trends that can be exploited to reveal previously unrecognized young stars in the solar neighborhood. To that end, we provide a series of plots of the EWs as functions of K-dwarf color and age: Figure 5 shows Na I, which can be used to evaluate surface gravity and age; Figure 6 includes H$\alpha$ and Ca II, both of which map activity; and Figure 7 shows Li I results that are most directly linked to age.

Figure 5(a) reveals subtle trends in the EW[Na I D] values when plotted against the colors of benchmark members in the various groups. There is a clear separation between the Hyades and field K dwarfs with isochrone ages compared to stars in the three younger moving groups. This gap is clearest for stars with $B_G$–$K$ = 2.5–3.5 but fades somewhat at the hotter and cooler ends of the K-dwarf temperature sequence. It is important to note that we observe no difference between Hyades members and K-dwarf field stars with age estimates over 1 Gyr. Figure 5(b) shows that no clear trend is observed when EW[Na I D] is plotted against estimated stellar age because the spread in temperatures within a group obfuscates the overall differences.

Figure 6(a) shows EW[H$\alpha$] strength versus color, with negative values indicating emission. There is a boundary between the young and old K dwarfs in the benchmark sample stretching through all temperatures, indicated by a broken black line at an EW value of 0.5 Å. In particular, the yellow squares represent generally old stars, including $o^2$ Eri, 20 Crt, and $\epsilon$ Indi that have ages estimated to be 3.7–4.6 Gyr. The EW[H$\alpha$] versus age comparison illustrated in Figure 6(b) shows the steady decrease of H$\alpha$ emission strength with increasing age for the five benchmark groups. K dwarfs in the younger associations, $\beta$ Pic, Tuc-Hor, and AB Dor, generally exhibit emission or filled-in line profiles; both cases are visible in the middle panel in the first row of Figure 2 for all three young moving groups. As stellar age increases, core emission fades, with all of the Hyades and field K dwarfs exhibiting H$\alpha$ absorption lines (Figure 3), corresponding to positive EW[H$\alpha$] values.

The bottom two panels in Figure 6 outline results for the other chromospheric activity diagnostic investigated, the Ca II line at 8452 Å. Figure 6(c) indicates that there is no clear trend for the stars of the benchmark sample when EW[Ca II] is plotted against color. A slight trend is seen when EW[Ca II] is plotted against age in Figure 6(d), with EW[Ca II] values generally increasing from younger to older groups. Careful examination of the Ca II line profiles presented in the fourth column of panels in Figures 2 and 3 indicates that for Hyades and K field stars there is no core re-emission, whereas effectively all stars in the younger associations show re-emission.

Results for the EW[Li I] diagnostic illustrated in Figure 7(a) indicate that all stars in $\beta$ Pic (blue points and line) show strong Li absorption signatures, regardless of $B_G$–$K$ color, while there are trends with color for members of both Tuc-Hor (green points and dotted line) and AB Dor (red points and dotted line). These trends can be seen visually in Figure 2, while Figure 3 shows that only a single star from the Hyades and K field groups, PX Vir, shows evidence of a Li I feature. Figure 7(b) illustrates the clear trend of Li I absorption strength and age. All nine K dwarfs observed in $\beta$ Pic exhibit strong Li I absorption, with EW[Li I] values of at least 0.25 Å. There is effectively no overlap between the $\beta$ Pic values and those of the slightly older Tuc-Hor and AB Dor members, two groups that show similar ranges in EW[Li I] values that are temperature dependent. As mentioned, all but one of the Hyades and field K dwarfs show no Li I signature, with values measured below what we consider to be our signal threshold of 0.05 Å in the CHIRON spectra. The exception is PX Vir, which has an estimated age of 300 Myr from Stanford-Moore et al. (2020) and showed an EW [Li I] value similar to K dwarfs in the Tuc-Hor and AB Dor groups.

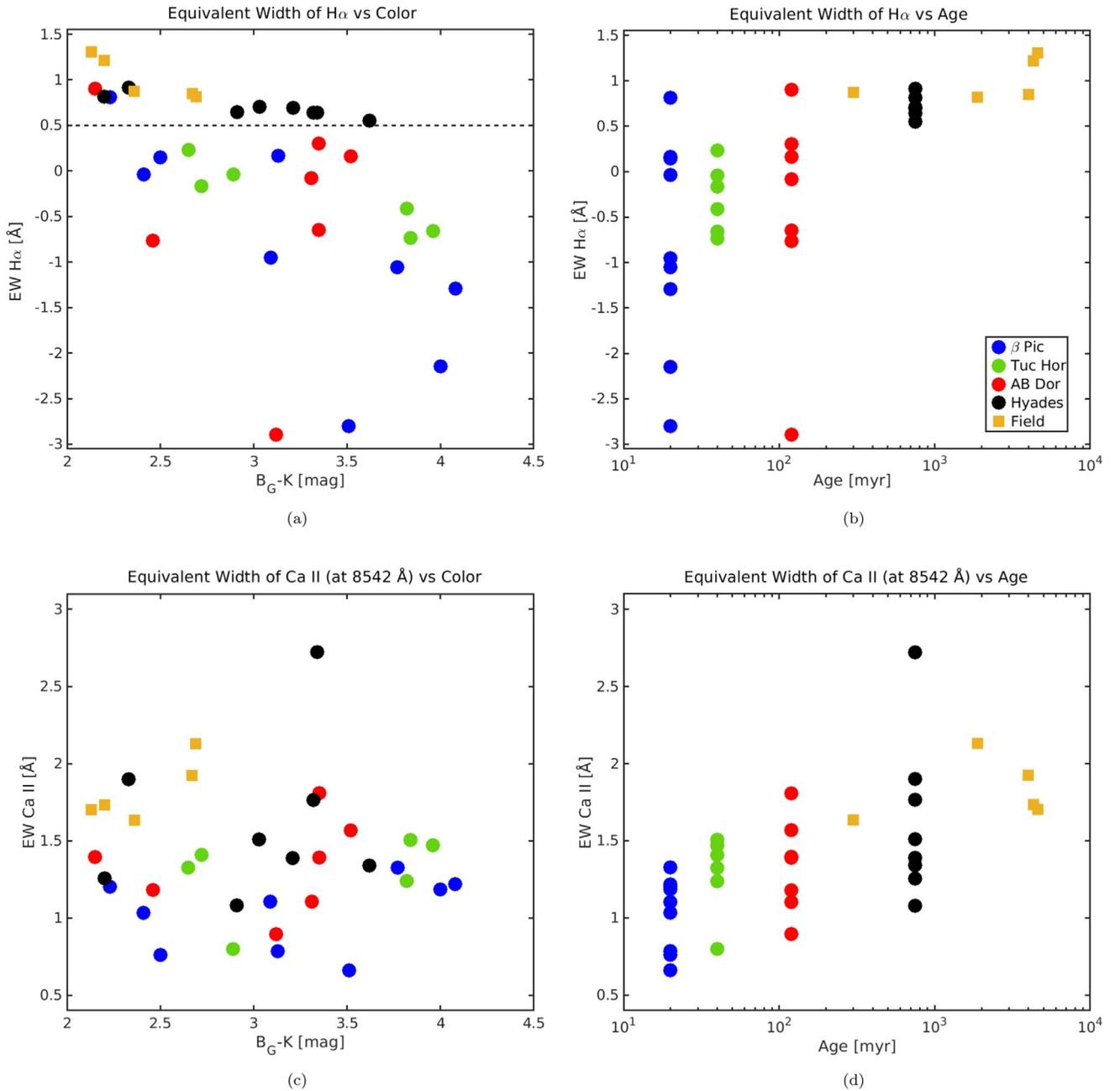

**Figure 6.** (a) Plot showing the EW[H$\alpha$] vs. $B_G$–$K$ color for the five groups of stars in the benchmark sample. The black dotted line separates the old and young stars. (b) Plot showing the EW[H$\alpha$] vs. estimated age for all five benchmark age groups. (c) Plot showing the EW[Ca II] vs. $B_G$–$K$ color for the five groups of stars in the benchmark sample. (d) Plot showing the EW[Ca II] vs. estimated age for all five benchmark age groups. Points are color-coded the same as in Figure 5.

### 5.2. Evaluation of Seven Radial Velocity Variable K Dwarfs

With trends in EW[Na I D], EW[H$\alpha$], and EW[Li I] established for groups in the benchmark sample, we can now evaluate the seven K dwarfs from our RVV sample using these three spectral diagnostics. The seven stars were chosen based on their variable RV measurements in early sets of CHIRON data. Their RV plots are presented in Figure 8, with a K field dwarf, HIP 042074, that shows no variable RV at the CHIRON precision resulting from our observing and reduction protocols, ~10 m s$^{-1}$. Figure 3 illustrates the regions around the spectral features for these seven stars.

Figure 9 illustrates the locations of the RVV stars, identified with circled numbers, overplotted on the results for the benchmark stars for the three relevant EWs individually and for the combination of EW[H$\alpha$] and EW[Li I]. Panel (a) shows where the RVV K dwarfs fall with respect to the EW[Na I D] versus $B_G - K$ trends that were established for the benchmark sample. Four (numbers 2 through 5) of the seven stars have EW [Na I D] values significantly below the Hyades trend line and with $B_G - K \approx 2.5 - 3.5$. These stars also stand out in panel (b) for H$\alpha$, with the same four K dwarfs falling below the activity boundary indicated by a dashed black line. In panel (c), the trend for EW[Li I] versus $B_G - K$ shows that two RVV stars (① and ②) have EW[Li I] greater than the 0.05 Å threshold set for a significant Li I signal. Both stars have $B_G - K < 3.0$, with ② V* V834 $\tau$ also standing out in panel (b) as active based on the H$\alpha$ trend.

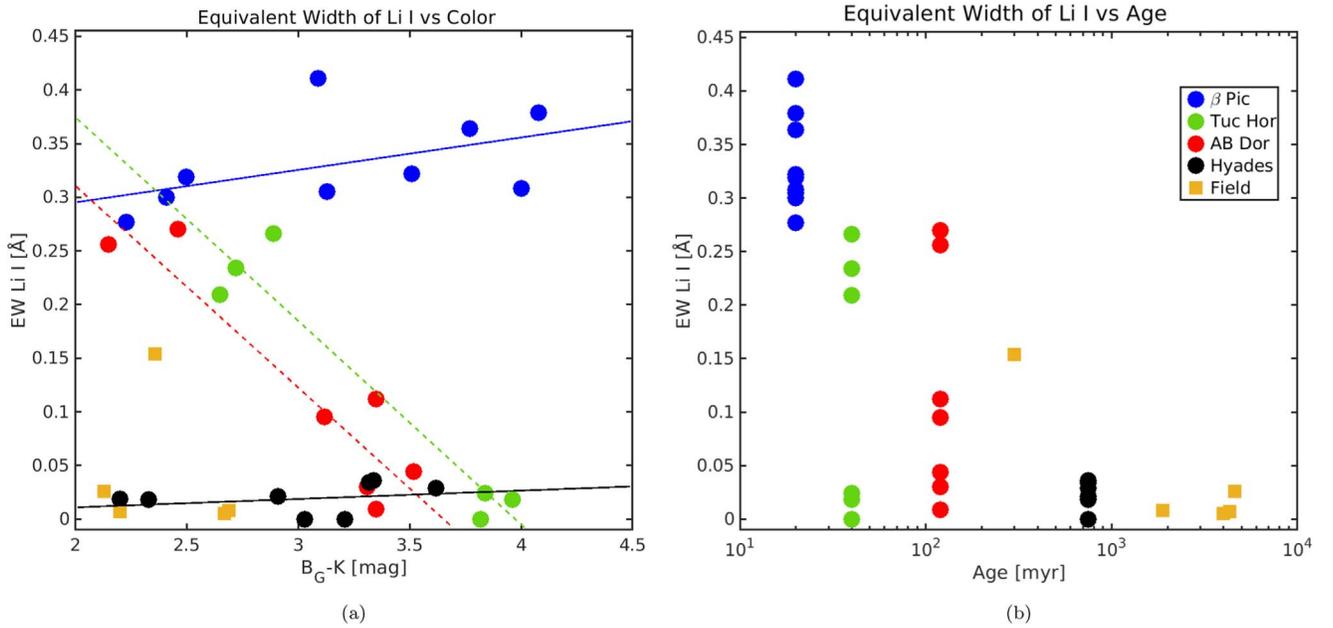

**Figure 7.** (a) Plot showing the EW[Li I] vs. $B_G-K$ color for the five groups of stars in the benchmark sample. Cluster trends are indicated by lines of the same color. (b) Plot showing the EW[Li I] vs. estimated age for all five benchmark age groups. Points are color-coded the same as in Figure 5.

Combinations of spectral features are often even more powerful than evaluations of individual lines, so in Figure 9(d) we show EW[H$\alpha$] versus EW[Li I] for both the benchmark sample stars and the seven RVV K dwarfs. There is a locus of stars representing the oldest benchmark groups in the lower left box of the plot, with borders set at EW[H$\alpha$] = 0.5 and EW[Li I] = 0.05. K dwarfs from the younger associations radiate from this mature locus, with stars in the youngest cluster, $\beta$ Pic, being the furthest removed and falling to the right of the plot. In addition to calibrating the methodology we will use on the full sample of more than 1200 stars, a key result of this initial effort is that we find five K dwarfs with variable RVs that land outside the old locus, with ① DX Leo being flagged as young based on Li I, ② V* V834 $\tau$ showing markers of both youth via Li I and activity via H$\alpha$, and the three stars ③ DG Cap, ④ BD +01 3657, and ⑤ HD 105065 flagged by H$\alpha$ as being active. We find that ⑥ HD 112099 and ⑦ BD +05 2529 show no signs of youth or activity, and their RV variations turn out to be caused by companions orbiting in periods of a few weeks to a few years, for which we confirm previously reported spectroscopic detections. In Section 6.3 we show that ③ DG Cap and ⑤ HD 105065 are newly identified spectroscopic binaries (type SB1) with close companions orbiting in less than 10 days and therefore could be young and/or have enhanced emission due to the close companions. In Section 6.3 we discuss all four spectroscopic binaries, and in Section 7 we provide additional details for all seven RVV stars.

### 5.3. Error Analysis

The EW[Li I] threshold mentioned above has been determined by comparing EW[Li I] measurements for the benchmark spectra to values published in other studies, namely, White et al. (2007) and López-Santiago et al. (2010). Comparing six K dwarfs, three from each study, to our CHIRON results, we find an average difference of 25 mÅ, with a range of differences from 3 to 30 mÅ. These comparisons increase our confidence in the EW[Li I] measurements made using our prescription (see Section 4.2), and we adopt a threshold of two times the average difference, 50 mÅ or 0.05 Å, based on these differences and to account for possible contamination by the unresolved Fe I line at 6707.44 Å. This contamination can be estimated using a prescription from Soderblom & Jones (1993), who give EW[Fe I 6707.441] = (20(B–V)–3) mÅ, which we estimate to be ∼16 mÅ for K dwarfs. Thus, our adopted 50 mÅ EW[Li I] threshold accounts for both the uncertainties in our measurement and a reasonable contribution of uncertainty due to Fe I 6707.44 Å contamination.

To estimate uncertainties in our EW[H$\alpha$] and EW[Ca II] measurements, we compared 7 of our 42 stars to White et al. (2007) and López-Santiago et al. (2010). Here we saw more variance between the previously published values and our own than for the EW[Li] values. We find that our measurements are typically consistent to ∼100 mÅ for each feature. The greater variations for these lines are not surprising given that both lines are linked to stellar activity that may vary over time, as well as the gaps in time between their observations and our own.

To test the repeatability of our own measurements, we examined multiple spectra of three K dwarfs of early, mid-, and late types, including five spectra for HIP 33817 with $B_p - R_p \lesssim 1.3$, nine spectra for HIP 70218 with $B_p - R_p = 1.3 - 1.6$, and five spectra for HIP 96285 with $B_p - R_p \gtrsim 1.6$. For EW[Na I], EW[H$\alpha$], and EW[Ca II], the variations were ∼30, ∼10, and ∼50 mÅ, respectively. Together, we estimate overall uncertainties for points in Figures 5 and 6, considering measurements by others and multiple measures of our own to be typically smaller than ∼150 mÅ.

### 5.4. Stellar Parameter Results

In addition to the EWs, Table 2 presents spectral parameter results for the sample. We report $T_{\rm eff}$, [Fe/H], and log $g$ values for 39 K dwarfs derived using Empirical SpecMatch. No values for the three stars with high rotation velocities—CD −64 1208, AB Dor, and LO Peg—are given because of high uncertainties. The $v \sin i$ values are also included, but because our $v \sin i$ values were derived from Empirical SpecMatch reference stars, they are limited to values $>10$ kms$^{-1}$, as the library of Yee et al. ([2017](#)) library sets this cutoff to ensure accurate spectral line matching.

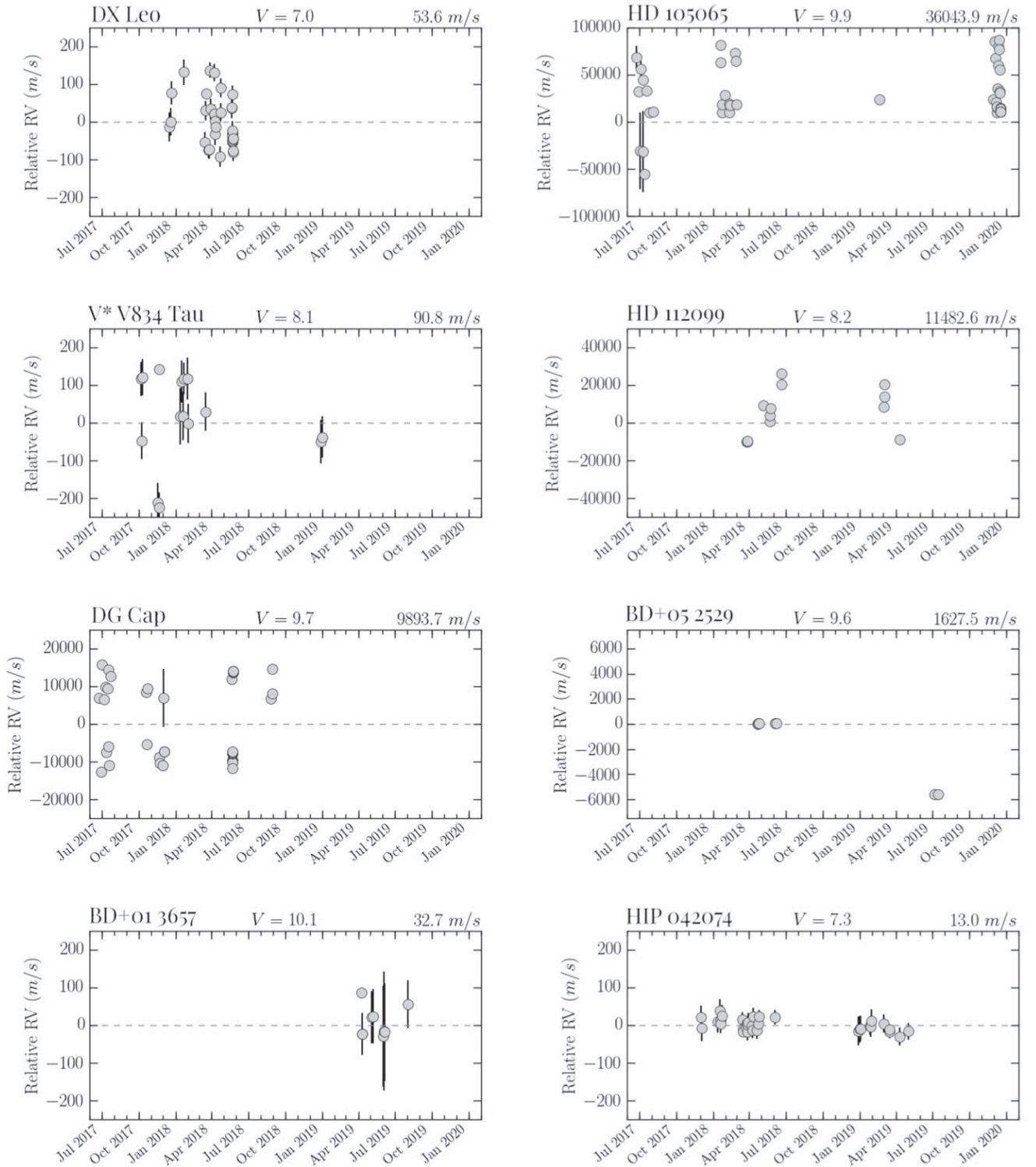

**Figure 8.** Left: plots showing RV results from CHIRON spectra for DX Leo (top row), V* V834 Tau (second row), DG Cap (third row), and BD +01 3657 (bottom row) for observations in 2017, 2018, and 2019. Right: plots showing RV results from CHIRON spectra for HD 105065 (top row), HD 112099 (second row), BD +05 2529 (third row), and HIP 042074 (bottom row, representing a typical K dwarf that is not variable in RV at CHIRON's precision) for observations in 2017, 2018, and 2019.

For members of the K field group, we observed negligible differences when comparing our derived $T_{\rm eff}$ values to existing measurements in the literature (Luck 2018; Aguilera-Gómez et al. 2018; Hojjatpanah et al. 2019). Four of the five field K dwarfs are in both papers, and we find that our $T_{\rm eff}$ measurements differ by only 20 K on average. As mentioned previously, the sequence of temperatures we measure exhibits a smooth correlation with $B_G - K$ color, as shown in Figure 4 (recall that the three points off the trend are the fast rotators), a further indication that our temperatures are reliable.

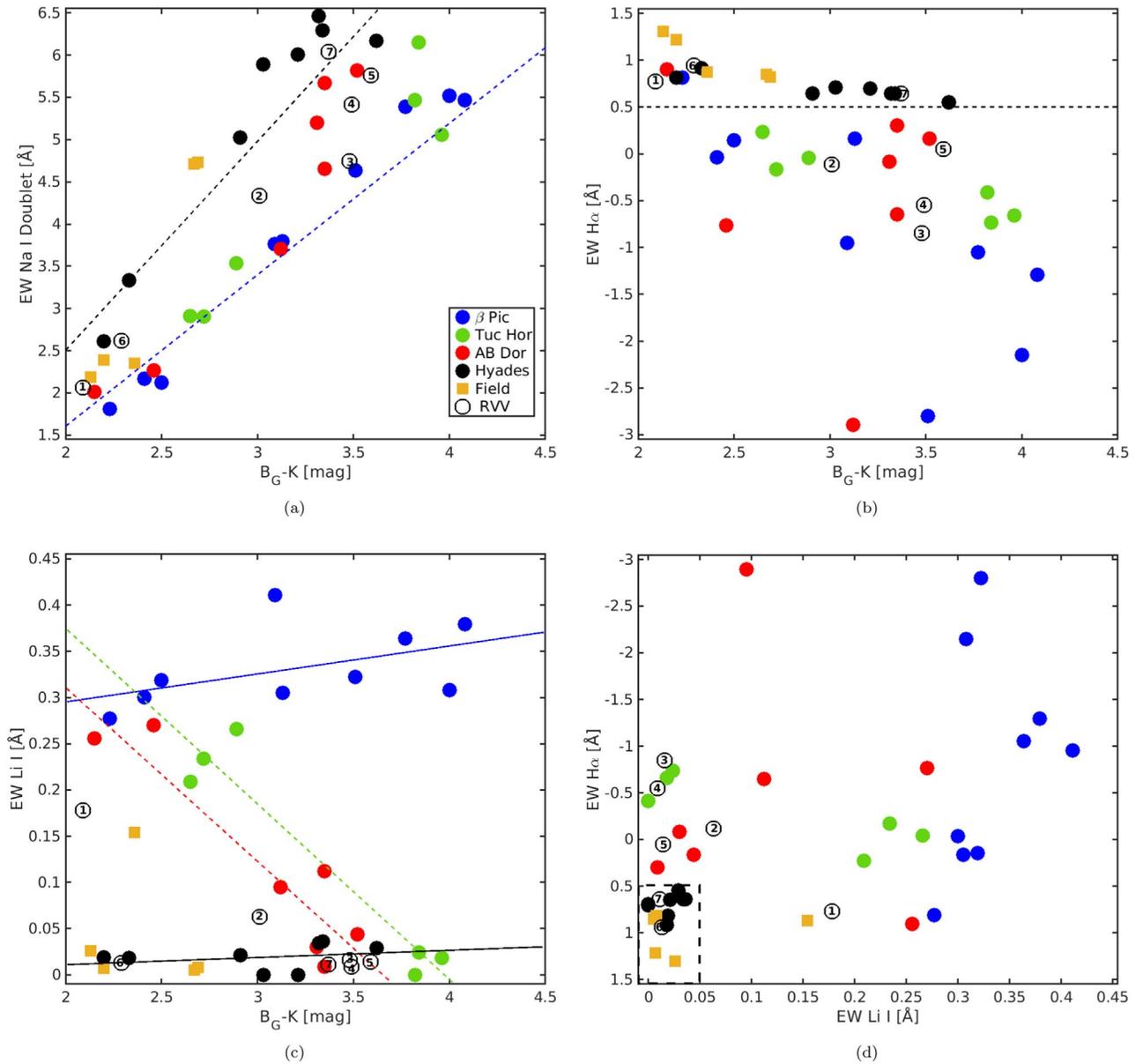

**Figure 9.** Plots placing the seven RVV stars (encircled numbers) among the set of 35 stars with estimated ages in the benchmark sample: (a) EW[Na I D] vs. $B_G - K$ color, (b) EW[H$\alpha$] vs. $B_G - K$ color, (c) EW[Li I] vs. $B_G - K$ color, and (d) EW[H$\alpha$] vs. EW[Li I], where the dashed black box in the bottom left encloses K dwarfs present in the old locus. Numbered stars: ① DX Leo, ② V* V834 $\tau$, ③ DG Cap, ④ BD +01 3657, ⑤ HD 105065, ⑥ HD 112099, and ⑦ BD +05 2529.

A comparison of derived [Fe/H] values for our stars to the Empirical SpecMatch library stars is shown in Figure 10. The [Fe/H] values derived for the four moving associations in our study were compared to values in Viana Almeida et al. (2009) for the $\beta$ Pic, Tuc-Hor, and AB Dor stars and to values in Perryman et al. (1998) for the Hyades members. We note that our ensemble averages for each group differed by no more than 30% from the values previously reported for these groups. Our [Fe/H] measurements of individual stars from the K field group agree with the reported values in Hojjatpanah et al. (2019) to 11%–17%, providing additional evidence that the values we derive with Empirical SpecMatch are consistent with previous determinations. We highlight the three relatively metal-poor stars with [Fe/H] $< -0.4$: $o^2$ Eri, 20 Crt, and DG Cap—none of the three is in a young group, and all appear to be reliable fits using Empirical SpecMatch.

Finally, log $g$ results were inconsistent for the younger moving groups, $\beta$ Pic and Tuc-Hor, but reliable for older stars in our sample. As shown in Figure 1, younger stars are clearly above the main sequence in the H-R diagram. However, Table 2 shows similar log $g$ values for K dwarfs of comparable $T_{\rm eff}$ even if they have markedly different ages. The likely reason for similar surface gravity values is that the Yee et al. (2017) spectral library includes primarily main-sequence stars,

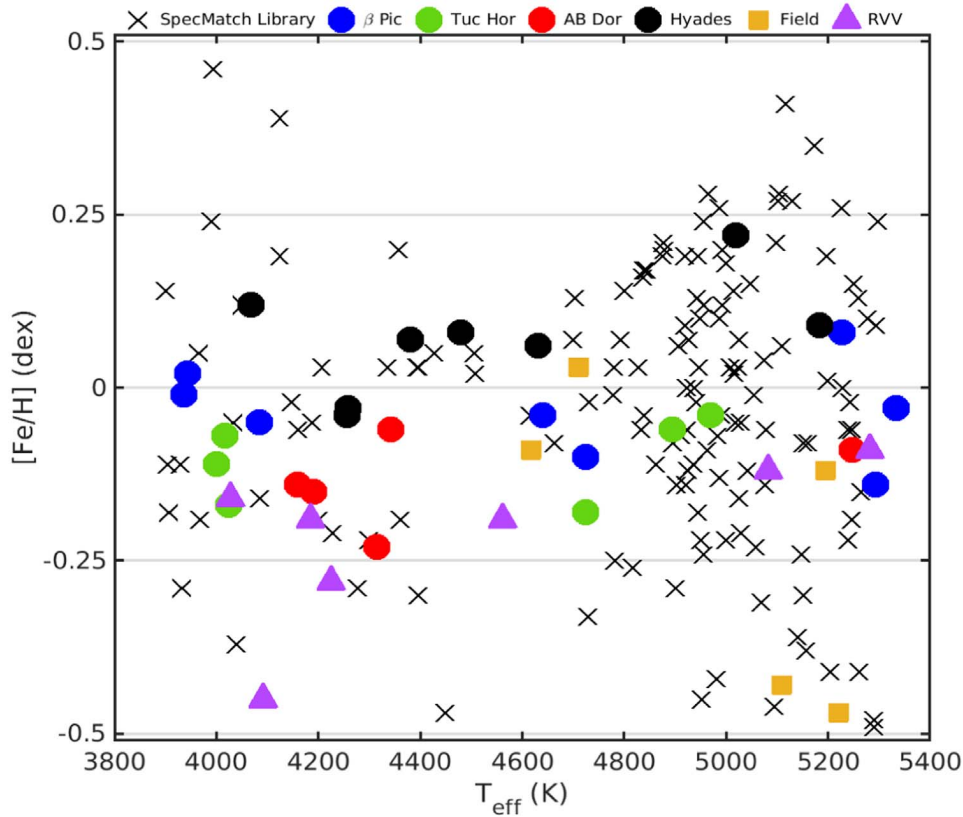

**Figure 10.** [Fe/H] plotted against effective temperature for 39 K dwarfs in our sample, with 143 stars (black crosses) from the Yee et al. (2017) SpecMatch library shown in the background.

so the log $g$ results for young stars reported here should be treated with caution.

### 5.5. γ Velocity and UVW Results

The UVW space motions for each K dwarf provide important information linking individual stars to the various moving groups and clusters and can indicate when those stars *do not* have matching motions in the Galaxy. Table 3 provides positions, proper motions, and parallaxes for the stars from Gaia, their $B_G$ magnitudes and $B_G - K$ colors, our derived γ systemic velocities from the CHIRON spectra and similar measurements from Gaia, and the derived UVW space velocities.

Figure 11 illustrates the dynamics of the stars in the samples. Panel (a) illustrates the comparison between the measured CHIRON γ velocities and those given in Gaia. Stars offset from the 1-to-1 line are multiples with orbits caught at different phases and/or are stars that are young, are active, or exhibit rapid rotation, any combination of which can make velocity measurements difficult. Outliers in the benchmark sample are labeled with encircled letters a through k. Stars with mismatched velocities are ⓐ HD 105065 (mildly active), ⓑ DG Cap (active), ⓒ LO Peg (rapid rotator), ⓓ HD 112099 (close multiple), ⓔ CD − 64 1208 (close multiple and rapid rotator), ⓕ HD 139084 (close multiple), ⓖ HD 161460 (close multiple), ⓗ PX Vir (close multiple), ⓘ AO Men (young, active), ⓙ AB Dor (close multiple and rapid rotator), and ⓚ HD 285828 (close multiple). Among the 11 outliers, seven are close multiples included in Table 4, which lists all stars in the sample known to have companions (more details in Section 6), indeed explaining the difference in the RV values for these stars.

Panels (b), (c), and (d) of Figure 11 illustrate the UVW space motions for stars in the benchmark samples. The ovals represent velocities for the four groups as reported in Riedel et al. (2014), where we have doubled the sizes of the value ranges given in order to trace the ellipses. There are six stars represented by encircled letters reported to be members of the groups that fall outside the ellipses. These are six out of seven moving group stars with RV values that are discordant between our measurements and those from Gaia. The implications of these offsets are that neither the CHIRON nor Gaia measurements may, in fact, represent each system's γ velocity, and only when an RV orbit is determined and the true systemic velocity derived can these stars be placed confidently in their relative groups.

Note that PX Vir has been identified as a member of the AB Dor moving group by Zuckerman et al. (2004). AB Dor group members typically have UVW space motions of (−7, −27, −13) (Riedel et al. 2014), but our derived measurements for PX Vir (−3, −31, −3) are significantly different from these values, given errors of less than 1 km s$^{-1}$. However, because PX Vir is a close binary, the offset in UVW may simply be due to measuring the close pair at an epoch with an RV that does not match the AB Dor moving group. Whereas PX Vir's EW [Li I] = 0.15 is consistent with AB Dor K dwarfs, it has EW [Hα] = 0.87, which is a deeper absorption feature than all but one AB Dor member, UY Pic. A final comparison can be made using the Ca II feature—UY Pic and all other slow rotators in the AB Dor group show a re-emission feature, but PX Vir does not. Given this ensemble of attributes, we conclude that PX Vir is *not* a member of AB Dor.

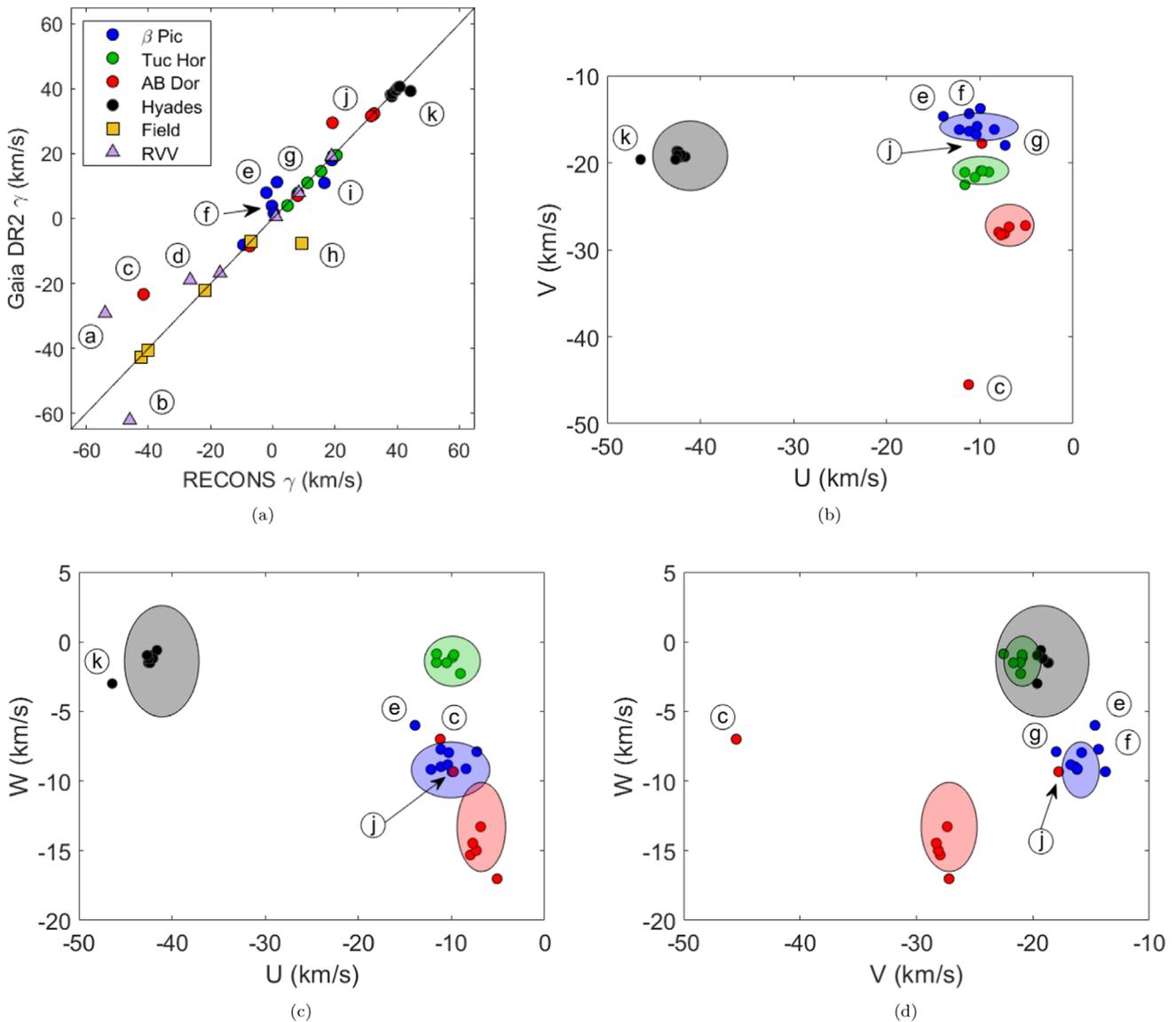

**Figure 11.** Diagrams illustrating the dynamics for members of the four moving groups and clusters used in the benchmark sample: (a) 1-to-1 correlation plot of CHIRON $\gamma$ velocity vs. Gaia $\gamma$ velocity, where the line represents equal values. (b, c, d) Galactic space motions $V$ vs. $U$, $W$ vs. $U$, and $W$ vs. $V$. Points represent members in $\beta$ Pic (blue circles), Tuc-Hor (green circles), AB Dor (red circles), Hyades (black circles), field K dwarfs (yellow squares), and RVV stars (purple triangles). Ovals represent characteristic $UVW$ velocities for the groups, and stars represented by encircled letters are outliers discussed in the text.

## 6. Companions

### 6.1. Sweep for Companions to All 42 Stars

We have used both Gaia EDR3 (Gaia Collaboration et al. 2021) and the Washington Double Star (WDS) catalog (Mason et al. 2001) to search for and confirm existing companions to the targets in our sample. In all, 16 of the 42 systems are multiples, with details given in Table 4. In addition to identifiers and coordinates, we list the parallaxes and errors from EDR3 and the separations ($\rho$) and position angles ($\theta$) from EDR3 or WDS. To assist in tracking the components, we also provide magnitudes, colors, and spectral types for each component, where available. Note that many potential components listed in WDS have been eliminated using Gaia data.

The search for companions in the EDR3 results was conducted using a search radius of $60'$, with an additional constraint of $\pi > 10$ mas, a value representative of the sample given that all of our sample stars are within 100 pc. This yielded a wealth of targets, which were then winnowed to potential companions worthy of scrutiny by imposing a condition that parallaxes needed to match within $\pm 0.5$ mas, a value that works well to identify true binaries when applied to stars within 100 pc during other RECONS-related surveys. However, for stars in moving groups it is possible that stars with similar $\pi$ values are not gravitationally bound companions, but merely members of the same moving association; this is especially evident in the Hyades cluster. To reveal true companions, we adopted a projected physical separation limit of $\leqslant 10^4$ au, a conservative value that captures the few widely separated true companions to stars in the solar neighborhood but allows for manageable vetting for possible companions that does not result in overwhelming numbers of false positives. We then consulted the WDS to determine whether or not the EDR3 companions were previously recorded. We also made a comprehensive sweep of all 42 stars in WDS for companions, including stars that may not have had any candidate companions

Table 4
K-dwarf Systems Confirmed via Gaia EDR3 and WDS to Have Companions

| Target WDS Name (1) | Comp (2) | R.A. (J2000.0) (3) | Decl. (J2000.0) (4) | $\pi$ (mas) (5) | $\sigma_\pi$ (mas) (6) | $\rho$ (arcsec) (7) | $\theta$ (deg) (8) | $B_G$ (mag) (9) | $K$ (mag) (10) | $B_G$–$K$ (mag) (11) | Spectral Type (12) |
|---|---|---|---|---|---|---|---|---|---|---|---|
| $o^2$ Eri | A | 04 15 16.3 | −07 39 10 | 199.608 | 0.121 | … | … | 4.61 | 2.50 | 2.11 | K0V |
| 04153-0739 STF518 | B | 04 15 21.8 | −07 39 29 | 199.691 | 0.051 | 83.34 | var[a] | 9.48 | 9.86 | −0.38 | DA3 |
| 04153-0739 STF518 | C | 04 15 21.5 | −07 39 20 | 199.452 | 0.069 | 78.10 | 97.5 | 11.49 | 5.96 | 5.53 | M5V |
| HD 285828[b] | A | 04 27 25.3 | +14 15 38 | 20.422 | 0.192 | … | … | 10.62J[c] | 7.71J | 2.91J | K2 |
| … | B | … | … | … | … | SB1[d] | … | … | … | … | … |
| HD 285830 | A | 04 27 47.0 | +14 25 04 | 20.353 | 0.016 | … | … | 9.68 | 7.35 | 2.33 | K0 |
| 04278+1425 PAT9 | B | … | … | … | … | 0.91 | 15.0 | … | … | … | … |
| AB Dor | A | 05 28 44.9 | −65 26 55 | 67.333 | 0.441 | … | … | 7.15 | 4.69 | 2.46 | K0V |
| AB Dor | C | … | … | … | … | 0.20 | ~156 | … | … | … | M8 |
| 05287-6527 CLO10 | Ba | 05 28 44.5 | −65 26 46 | 66.992 | 0.081 | 8.92 | 347.1 | 13.08J | 7.34J | 5.74J | M5 |
| 05287-6527 CLO10 | Bb | … | … | … | … | ⩽2.0 | var | … | … | … | L5 |
| UY Pic | A | 05 36 56.9 | −47 57 53 | 40.657 | 0.014 | … | … | 7.96 | 5.81 | 2.15 | K0V |
| 05369-4758 HDS751 | B | 05 36 55.1 | −47 57 48 | 40.758 | 0.040 | 18.26 | 285.9 | 10.02 | 6.61 | 3.42 | K6Ve |
| HD 50281 | A | 06 52 18.1 | −05 10 25 | 114.355 | 0.042 | … | … | 6.80 | 4.11 | 2.69 | K3V |
| 06523-0510 WNO17 | B | 06 52 18.0 | −05 11 24 | 114.291 | 0.022 | 58.83 | 180.6 | 10.33 | 5.72 | 4.60 | M2V |
| DX Leo | A | 09 32 43.8 | +26 59 19 | 55.329 | 0.021 | … | … | 7.21 | 5.12 | 2.09 | K0V |
| 09327+2659 LDS3903 | B | 09 32 48.2 | +26 59 44 | 55.292 | 0.071 | 65.01 | 67.3 | 16.06 | 9.47 | 6.58 | M5V |
| 20 Crt | A | 11 34 29.5 | −32 49 53 | 104.613 | 0.028 | … | … | 6.15 | 4.14 | 2.02 | K0V |
| 11345-3250 LDS6245 | B | 11 34 30.5 | −32 50 02 | 104.657 | 0.027 | 15.30 | 128.3 | 14.92 | … | … | DC8 |
| BD +05 2529[e] | A | 11 41 49.6 | +05 08 26 | 31.762 | 0.381 | … | … | 9.80J | 6.43J | 3.37J | K4V |
| … | B | … | … | … | … | SB1 | … | … | … | … | … |
| HD 105065 | A | 12 05 50.7 | −18 52 31 | 43.365 | 0.092 | … | … | 10.21 | 6.62 | 3.59 | K5V |
| 12058-1853 WNO54 | B | 12 05 46.6 | −18 49 32 | 43.695 | 0.030 | 187.77 | 342.2 | 16.49 | 10.32 | 6.18 | M5V |
| HD 112099[f] | A | 12 53 54.4 | +06 45 46 | 36.594 | 0.061 | … | … | 8.44J | 6.15J | 2.29J | K1V |
| … | B | … | … | … | … | SB1 | … | … | … | … | … |
| PX Vir | A | 13 03 49.7 | −05 09 43 | 46.100 | 0.810 | … | … | … | 5.72 | … | K1V |
| 13038-0510 EVT3 | B | … | … | … | … | <0.05 | var | … | 7.00 | … | … |
| HD 139084 | Aa | 15 38 57.6 | −57 42 27 | 25.829 | 0.201 | … | … | 8.08J | 5.85J | 2.23J | K0V |
| 15390-5742 NLS2 | Ab | … | … | … | … | ~0.1 | var | … | … | … | MV |
| 15390-5742 SKF1501 | B | 15 38 56.8 | −57 42 19 | 25.440 | 0.017 | 10.32 | 323.3 | 14.95 | 9.19 | 5.77 | M5Ve |
| HD 161460 | A | 17 48 33.7 | −53 06 43 | 13.053 | 0.114 | … | … | 9.28J | 6.78J | 2.50J | K0IVJ |
| 17486-5307 CVN29 | B | … | … | … | … | SB2 | … | … | … | … | … |
| … | C[g] | 17 48 33.7 | −53 06 12 | 13.049 | 0.019 | 31.5 | 0.1 | 14.21 | 9.27 | 4.95 | … |
| 18454-6452 SKF105 | A | 18 45 26.9 | −64 52 17 | 34.736 | 0.158 | … | … | 4.83 | 4.30 | 0.53 | A7V |
| CD −64 1208 | Ba | 18 45 36.9 | −64 51 47 | 35.158 | 0.181 | 71.34 | 64.6 | 9.61 | 6.10 | 3.51 | K5Ve |
| 18454-6452 BIL4 | Bb | … | … | … | … | 0.18 | ~95 | … | … | … | K7 |
| $\epsilon$ Indi | A | 22 03 21.7 | −56 47 10 | 274.843 | 0.096 | … | … | 4.91 | 2.24 | 2.67 | K4V |
| 22034-5647 SOZ1 | Ba | 22 04 10.5 | −56 46 57 | 270.658 | 0.690 | 402.46 | 88.4 | 21.14J | 11.21J | 9.93J | T1V |
| 22034-5647 VLK1 | Bb | … | … | … | … | <1.0 | var | … | … | … | T6V |

**Notes.** Column (1): object identifier (top) and WDS designation (bottom). Column (2): system component. Column (3): right ascension. Column (4): declination. Column (5): trigonometric parallax. Column (6): error in trigonometric parallax. Column (7): angular separation from primary. Column (8): position angle relative to primary, where north is 0° and east is 90°. Column (9): Gaia EDR3 BP magnitude = $B_G$. Column (10): 2MASS $K$ magnitude. Column (11): color. Column (12): spectral type. Names are provided for targets included in the target sample and are not necessarily the system's primary.
[a] Varies in value.
[b] Binary determination provided by Griffin (2012).
[c] Joint value indicated with J includes components below it.
[d] SB1/SB2 indicates a single-lined/double-lined spectroscopic binary.
[e] Binary determination provided by Sperauskas et al. (2019).
[f] Binary determination provided by Griffin (2009).
[g] New tertiary (see Section 6.2).

in EDR3. We found that the companions with $\rho > 2''$ were generally in both EDR3 and WDS and those at smaller separations were in WDS only.

### 6.2. Discovery of a Tertiary in the HD 161460 System

During the companion search, we found a previously unreported companion to the known binary, HD 161460, which is a member of $\beta$ Pic. The third component is a 14th mag star separated 31.5″ from the primary, corresponding to a projected separation of ~2400 au, and is found at a position angle of 0°.1, effectively due north. This was not found to be a companion candidate in our EDR3 search because astrometric information for the primary is absent, presumably because it is a close binary. However, Gaia DR2 parallax measurements confirm the companion to be real, given parallaxes in DR2 differing by only 0.004 mas. Matching proper motions between both components provide further verification that the tertiary is part of the HD 161460 system.

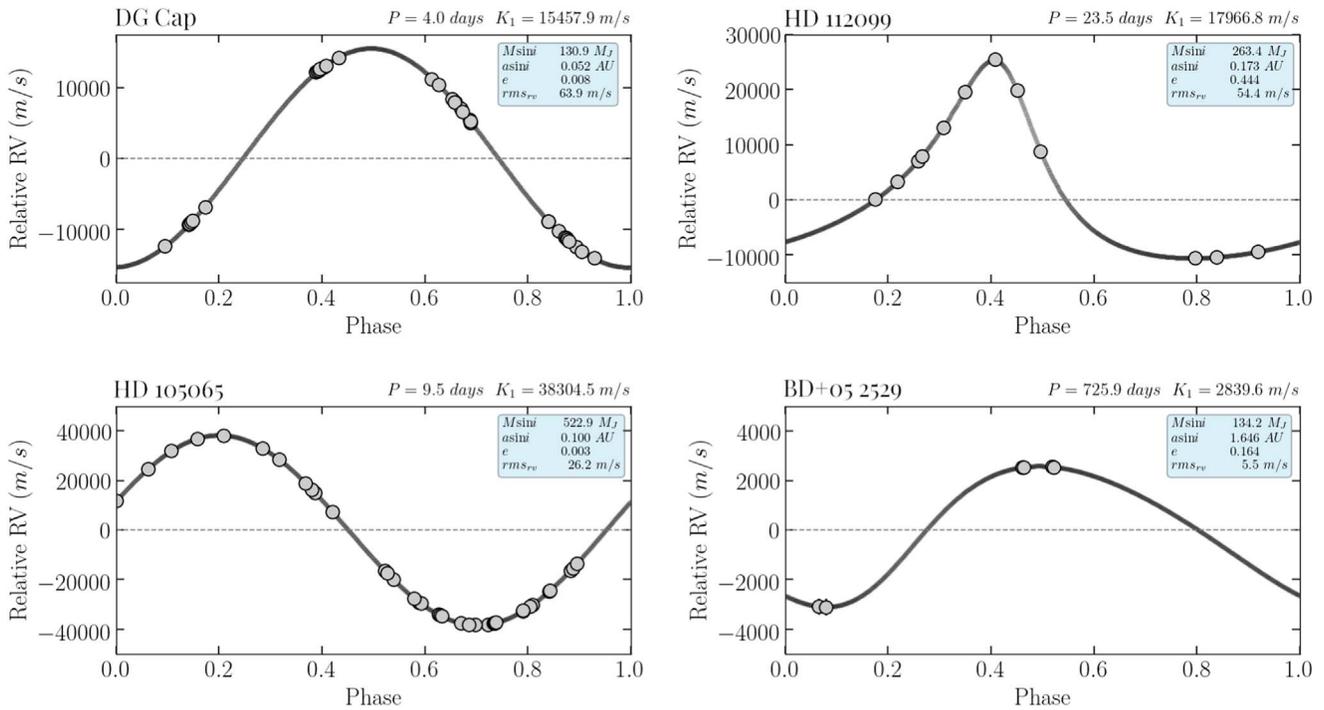

**Figure 12.** Left: phase-folded RV curves for the newly discovered binaries DG Cap (top) and HD 105065 (bottom). Right: phase-folded RV curves for the previously reported binaries HD 112099 (top) and BD +05 2529 (bottom).

*6.3. Spectroscopic Binaries in Our Variable Radial Velocity Sample*

Of the seven K dwarfs targeted as RVV stars, four turned out to be spectroscopic binaries for which insufficient CHIRON data were available for solutions when this sample was created. The companions of DG Cap (HIP 102119) and HD 105065 (HIP 59000) are reported here for the first time, while those of HD 112099 (HIP 62942) and BD +05 2529 (HIP 57058) were previously reported by Griffin (2009) and Sperauskas et al. (2019), respectively. CHIRON data were reduced as described in Section 3 and phased to produce the RV curves shown in Figure 12 for DG Cap (HIP 102119) with a period of 4.0 days and HD 105065 (HIP059000) with a period of 9.5 days. In Section 5.1 both ③ DG Cap and ⑤ HD 105065 were flagged as being active based on the presence of H$\alpha$ emission that is shown in Figure 3, but the activity may be induced by the close companion in each case.

Figure 12 also shows the CHIRON orbits of HD 112099 (HIP062942) and BD +05 2529 (HIP057058). For HD 112099, the 13 CHIRON measurements taken over 12 months yield an orbital period of 23.5 days, matching the value given by Griffin (2009). For BD +05 2529, only eight CHIRON measurements are available over 15 months, but that is sufficient to produce an orbital fit with a period of 726 days that is consistent with that derived by Sperauskas et al. (2019). While more data should be acquired to firm up both orbits, it is clear that the CHIRON data indicate changing RVs due to companions in both cases, and that neither system is necessarily young.

### 7. Systems Worthy of Note

Here we provide details for systems with notable attributes, listed in order of R.A., with coordinate designations to start each entry. This list includes all of the RVV stars (numbered ① through ⑦) in Figure 9 and the lettered stars worthy of further comment in Figure 11(a).

0427+1415: ⓚ HD 285828 is a known SB1 (Griffin 2012), which provides a ~5 km s$^{-1}$ offset in RV measurements between our CHIRON and Gaia epochs. This results in a slight offset of the point from the Hyades ovals in the panels of Figure 11. Nonetheless, the star is, indeed, a member of the Hyades.

0441+2054: ② V* V834 $\tau$ is a known rapid rotator and was classified as a BY Draconis variable by Mishenina et al. (2012). BY Draconis variables are K- or M-type main-sequence stars that exhibit variations in luminosity over short time periods primarily owing to excessive rotation (Mishenina et al. 2012). Mamajek & Hillenbrand (2008) reported an estimated age for V* V834 $\tau$ of 49 ± 37 Myr, using rotation–age and calcium H and K emission–age correlations. This age estimate aligns with the trends shown in Figure 9, as all four plots presented there identify V* V834 $\tau$ as young and active. Lehtinen et al. (2016) list this dwarf as a member of the Ursa Major moving group. However, using BANYAN $\Sigma$, we find V* V834 $\tau$ to be a field dwarf.

0528–6526: ① AB Dor is the eponymous quadruple system central to the AB Doradus moving group. It is composed of two wide binaries with primaries known as A and B separated by ~9″, each with its own close component. As noted by Zuckerman et al. (2004), the AB Dor components have rapid rotation rates, with the A component rotating in 0.514 days and the B component in 0.38 days (Azulay et al. 2015). This elucidates both its highly broadened spectral lines displayed in Figure 2 and the difference between its $\gamma$ velocity as measured by us and Gaia (Table 3). Our $\gamma$ velocity for component A was used to compute the system's $UVW$, but that value catches the close pair at a velocity that does not represent the system's true velocity. Therefore, AB Dor falls outside the AB Dor region in Figure 11.

0932+2659: ① DX Leo presented the greatest EW[Li I] value, 0.18 Å, of all the RVV K dwarfs examined. The star has $B_G - K = 2.09$, which meets our K-dwarf inclusion criteria despite the G9V spectral type given in SIMBAD. DX Leo is classified as a BY Draconis variable in SIMBAD owing to observed changes in luminosity caused by rapid rotation (Griffin 1994). Using the Gaia parallax and position data, along with the Gagne et al. (2018) BANYAN $\Sigma$ code, we were able to show that DX Leo is a field star and not a member of any known moving association. Wang & Wei (2009) had already identified DX Leo as a young dwarf with an EW[Li I] value of $0.20 \pm 0.005$ Å and an age estimate of 30–100 Myr. We note that both results are similar to values measured for K dwarfs in the Pleiades, which have EW[Li I] of 0.1–0.2 Å and are ~125 Myr old (Bouvier et al. 2018). Thus, our findings corroborate this age estimation and firmly characterize DX Leo as a young K dwarf within 20 pc.

1141+0508: ⑦ BD +05 2529 is listed as a double-lined spectroscopic binary (SB2) by Halbwachs et al. (2018) using CORAVEL spectra. A year later, Sperauskas et al. (2019) used CORAVEL plus VUES echelle spectrograph data to identify BD +05 2529 as an SB1. In our study that was solely dependent on CHIRON spectra, we observed no evidence of double lines for this spectroscopic binary. However, in Figure 1 we clearly observe BD +05 2529 as the RVV star that is elevated above the main sequence, presumably because of the excess flux of the companion. There is an absence of the Li I, H$\alpha$, and Ca II signatures, as illustrated in Figure 3 and panels (b) and (c) of Figure 9. BANYAN $\Sigma$ identified this as a field dwarf, and we concur.

1205–1852: ⑤ HD 105065 (also ⓐ) was identified as active by Martínez-Arnáiz et al. (2010) with a $\log R'_{HK}$ of −4.12 and by Gray et al. (2006) with an S-index of 3.15. Using the $\log R'_{HK}$ thresholds of activity set by Henry et al. (1996), HD 105065 is classified as very active with a value $> -4.20$. H$\alpha$ is in absorption (Table 2), but the feature is weaker than for Hyades cluster K dwarfs with estimated ages ≥700 Myr, implying youth. However, we have found that this is a close binary with orbital period of 9.5 days, and the companion may be inducing the activity. BANYAN $\Sigma$ indicates that this is a field dwarf.

1253+0645: For ⑥ HD 112099 (also ⓐ), our study reported EW[H$\alpha$] and EW[Li I] values below the thresholds set for high activity and youth. Panel (d) of Figure 9 illustrates this result with HD 112099 located firmly in the old locus of K dwarfs. Table 4 shows that Strassmeier et al. (2000) identified this target as a single-lined spectroscopic binary (SB1); however, no orbital parameters were included with the first detection. Later, Griffin (2009) used observations from Strassmeier et al. (2000) to estimate a period of 23.5 days for the binary system. In contrast to our findings of inactivity based on the lack of H$\alpha$ and Ca II emission in Figure 3, Gray et al. (2003) reported a $\log R'_{HK}$ of −4.71 and classified the star as active. However, using the Henry et al. (1996) $\log R'_{HK}$ thresholds, HD 112099 would only be described as moderately active. BANYAN $\Sigma$ indicates that this is a field dwarf, and we conclude that it is not particularly young.

1303–0509: ⓑ PX Vir is a close binary that has been previously identified as a member of the AB Doradus moving group (Schaefer et al. 2018; Zuckerman et al. 2004). However, we find from its spectral features that it shows signs of both inactivity and youth. The spectrum of PX Vir shown in Figure 3 contains H$\alpha$, Ca II (with no core emission), and Li I features, all in absorption. The former two indicate the system's lack of activity, which would typically coincide with older stars. Yet this is the only member of the field group to exhibit clear Li I absorption, a signature of youth. In addition, BANYAN $\Sigma$ indicates that PX Vir is not a member of AB Dor and is instead a field star. We propose that the PX Vir system be reclassified to the group of field stars.

1822+0142: ④ BD +01 3657 is listed in SIMBAD as a rotationally variable star. Gray et al. (2006) tagged the dwarf as very active, with a −4.055 dex $\log R'_{HK}$ value, which is derived from the chromospherically sensitive calcium H and K lines. We observe strong core emission in H$\alpha$, as shown in Figure 3, and derive EW[H$\alpha$] = −0.55, which agrees with the Gray et al. (2006) result. BANYAN $\Sigma$ identified BD +01 3657 as a field dwarf, but it shows clear evidence of activity, implying youth.

1845–6451: ⓒ CD −64 1208 is a close multiple. This triple exhibits some of the most rapid rotation observed in the $\beta$ Pic moving group (García-Alvarez et al. 2011), with a reported period of 0.354 days by Messina et al. (2010). This explains the system's nearly spectral features (Figure 2) and disagreement between our $\gamma$ velocity and that from Gaia.

2041–2219: ③ DG Cap (also ⓑ) is listed as having a very active chromosphere by Gray et al. (2006). Our chromospheric indicators, H$\alpha$ and Ca II, confirm the star to be active, with significant core emission in both lines (Figure 3)—it exhibits the strongest H$\alpha$ emission of the seven RVV stars, with EW[H$\alpha$] = −0.85. This star turns out to be the fastest orbiting spectroscopic binary of the seven RVV stars, with an orbital period of only 4.0 days. Presumably the companion is responsible for inducing the activity, given that the BANYAN $\Sigma$ code showed no membership to any known moving associations.

2131+2320: ⓒ LO Peg is a variable star with an active chromosphere and an ultrafast rotational period of 0.423 days (Jeffries et al. 1994; García-Alvarez et al. 2011). We show its Doppler-broadened spectrum in Figure 2, which presumably explains the difference between our $\gamma$ velocity and that from Gaia.

## 8. Conclusions

Here we summarize the major outcomes of this research:

1. We have created a reliable spectroscopic methodology that separates K dwarfs of various ages and activity levels using a benchmark sample of 35 stars from four moving groups with known ages, augmented with a handful of the nearest K dwarfs with ages from gyrochronology. This methodology can be applied to field K dwarfs to extract the small fraction of K dwarfs younger than ~145 Myr. In a first effort, we evaluate seven field stars that exhibit high RV variability in CHIRON data, selected from 300 nearby K dwarfs that have multiple observations.

2. Using high-resolution spectra acquired with the CHIRON spectrograph on the SMARTS 1.5 m and Empirical SpecMatch, we derive system $\gamma$ velocities, $T_{\rm eff}$, [Fe/H], log $g$, $v \sin i$, and Na I, H$\alpha$, Li I, and Ca II equivalent line widths for the full sample of 42 K dwarfs.

3. $\gamma$ velocity variations in CHIRON data were used to identify stars that are potentially young and/or active, or that are close multiples. *UVW* space motions were derived and used to confirm or refute K-dwarf membership to specific moving groups. PX Vir shows a Li I absorption feature but was dismissed from the AB Dor moving group based on its *UVW* space motions and the absence of a Ca II activity marker (Figure 2).

4. We have confirmed the known young stars ① DX Leo and ② V* V834 $\tau$ using our benchmark sample and the chosen age diagnostic Li I. Space motions for both stars do not match any of the young moving groups discussed here.

5. Using the H$\alpha$ diagnostic, we have confirmed the activity of ③ DG Cap, ④ BD +01 3657, and ⑤ HD 105065. This indicates that ④ BD +01 3657 is possibly young, while ③ DG Cap and ⑤ HD 105065 are likely active because of interactions with newly discovered close companions with orbital periods of 4.0 and 9.5 days, respectively.

6. Two K dwarfs exhibiting RV variations, ⑥ HD 112099 and ⑦ BD +05 2529, have been shown to lack the spectral markers of age and activity used in this study. Both have been confirmed to be spectroscopic binaries with CHIRON data, supporting other studies (Strassmeier et al. 2000; Halbwachs et al. 2018).

## ORCID iDs

Hodari-Sadiki Hubbard-James https://orcid.org/0000-0003-4568-2079
Todd J. Henry https://orcid.org/0000-0002-9061-2865
Leonardo A. Paredes https://orcid.org/0000-0003-1324-0495
Azmain H. Nisak https://orcid.org/0000-0002-1457-1467